\documentclass[12pt,english]{article}
\usepackage{lmodern}
\usepackage{lmodern}
\usepackage[T1]{fontenc}
\usepackage[cp1252]{inputenc}
\usepackage{geometry}
\usepackage{color}
\usepackage{babel}
\usepackage{amsmath}
\usepackage{amsthm}
\usepackage{amssymb}
\usepackage{setspace}
\usepackage[authoryear,comma]{natbib}
\usepackage{microtype}
\usepackage[unicode=true,
 bookmarks=false,
 breaklinks=false,pdfborder={0 0 1},backref=section,colorlinks=true]
 {hyperref}
\hypersetup{pdftitle={"RAUM"},
 pdfauthor={"KashaevAguiar"},
 pdfnewwindow=true,pdfstartview=FitH,urlcolor=blue!90!red!45!black,citecolor=blue!90!red!45!black,linkcolor=red!90!black}

\makeatletter
\theoremstyle{plain}
\newtheorem{thm}{\protect\theoremname}
\theoremstyle{definition}

\theoremstyle{definition}
\newtheorem{example}[thm]{\protect\examplename}
\theoremstyle{plain}

\usepackage{amsfonts}
\usepackage{dsfont}\usepackage{mathrsfs}\usepackage{ushort}

\usepackage{titlesec}\usepackage{titling}
\usepackage{caption}
\usepackage{enumitem}\usepackage{booktabs}
\usepackage{tikz}
\usepackage{pstricks}\usepackage{pst-all}
\usepackage{pst-plot}
\usepackage{pst-node}\usepackage{pst-3dplot}\usepackage{sgamevar}
\usepackage{subcaption}
\usepackage{lscape}
\usepackage{pifont}
\usepackage{longtable}
\usepackage{algorithmic}
\usepackage{pdflscape}
\usepackage{rotating}
\usepackage{float}

\setenumerate{label=\small(\roman*)}
\DeclareCaptionFont{fancy}{\bfseries\sffamily}
\gamemathtrue
\allowdisplaybreaks

\providecommand{\psreset}{\psset{%
		linewidth=0.3pt,linestyle=solid,linecolor=black,
		dotsize=2.5pt,dotsep=2.5pt,arrowsize=4pt,
		fillstyle=none,fillcolor=white,
		showpoints=false,arrows=-,linearc=0,framearc=0,
		hatchsep=2pt,hatchwidth=0.2pt,nodesep=4pt,opacity=1}
	\psset{gridcolor=black!60, subgridcolor=black!30}
}

\psreset

\usepackage{graphicx}
\graphicspath{ {figures/} }

\titleformat{\section}[block]{\centering\large\bfseries\sffamily}{\thesection.}{0.5em}{}
\titleformat{\subsection}[block]{\flushleft\bfseries}{\thesubsection.}{0.5em}{}
\titleformat{\subsection}[block]{\flushleft\bfseries\sffamily}{\thesubsection.}{0.5em}{}
\titleformat{\subsubsection}[runin]{\normalsize\bfseries\sffamily}{\bfseries\upshape\sffamily\thesubsubsection.}{0.5em}{}[.--\:]
\renewcommand{\thesubsubsection}{\arabic{section}.\arabic{subsection}.\arabic{subsubsection}}
\titlespacing{\section}{0ex}{10ex}{5ex}
\titlespacing{\subsection}{0in}{6ex}{3ex}
\titlespacing{\subsubsection}{0mm}{2ex}{0.5em}
\pretitle{\begin{center}\LARGE\bfseries\sffamily}
\posttitle{\par\end{center}\vskip 0.5em}
\preauthor{\begin{center} \large \lineskip 0.5em\begin{tabular}[t]{c}}
\postauthor{\end{tabular}\par\end{center}}
\predate{\begin{center}\small}
\postdate{\par\end{center}}
\providecommand{\abstitle}[1]{{\par\vspace*{2ex}\small\bfseries\sffamily #1}\hspace*{1ex}}
\renewenvironment{abstract}%
{\begin{center}\begin{minipage}{0.8\linewidth}%
			\abstitle{Abstract}\small}%
		{\end{minipage}\end{center}\vfill\clearpage}

\providecommand{\Char}[1]{\mathds{1}\left(\,#1\,\right)}
\providecommand{\Real}{{\mathds{R}}}

\providecommand{\tr}{^{\prime}}

\providecommand{\rand}[1]{\mathbf{#1}}
\providecommand{\rands}[1]{\boldsymbol{#1}}

\providecommand{\abs}[1]{\left\lvert#1\right\rvert}

\theoremstyle{remark}
  \theoremstyle{plain}
  \newtheorem{lemma}{\protect\lemmaname}\theoremstyle{definition}
    \newtheorem{proposition}{\protect\propositionname}\theoremstyle{definition}
  \newtheorem{definition}{\protect\definitionname}\theoremstyle{plain}
\newtheorem{theorem}{\protect\theoremname}\theoremstyle{plain}
  \theoremstyle{definition}
  \newtheorem{assumption}{\protect\assumptionname}%
  \providecommand{\assumptionname}{Assumption}

  \providecommand{\definitionname}{Definition}
  \providecommand{\lemmaname}{Lemma}
  \providecommand{\propositionname}{Proposition}
  \providecommand{\remarkname}{Remark}
\providecommand{\corollaryname}{Corollary}
\providecommand{\theoremname}{Theorem}
\providecommand{\examplename}{Example}

\makeatother

\providecommand{\definitionname}{Definition}
\providecommand{\examplename}{Example}
\providecommand{\lemmaname}{Lemma}
\providecommand{\theoremname}{Theorem}

\begin{document}
\title{A Random Attention and Utility Model\thanks{The ``\textcircled{r}'' symbol indicates that the authors' names are in certified random order, as described by \citet{ray2018certified}. We thank the editor, the associate editor, and two anonymous referees for excellent suggestions that have greatly improved the manuscript. We gratefully acknowledge financial support from the Western Social Science Faculty Grant (FRDF R5533A02) and Social Sciences and Humanities Research Council Insight Development Grant.}}

\author{ 
	Nail Kashaev \textcircled{r}
	Victor H. Aguiar\thanks{Kashaev: Department of Economics, University of Western Ontario; \href{mailto:nkashaev@uwo.ca}{nkashaev@uwo.ca}. Aguiar: Department of Economics, University of Western Ontario; \href{mailto:vaguiar@uwo.ca}{vaguiar@uwo.ca}.}
	}
\date{This version: May, 2022 / First Version: January, 2021}

\maketitle

\begin{abstract}
We generalize the stochastic revealed preference methodology of \citet{mcfadden1990stochastic} for finite choice sets to settings with limited consideration. Our approach is nonparametric and requires partial choice set variation. We impose a monotonicity condition on attention first proposed by \citet{cattaneo2017random} and a stability condition on the marginal distribution of preferences. Our framework is amenable to statistical testing.  These new restrictions extend widely known parametric models of consideration with heterogeneous preferences.  


\end{abstract}

\section{Introduction}
The stochastic revealed preference methodology of \citet{mcfadden1990stochastic} is a cornerstone of economic analysis. This research agenda aims at explaining the behavior of a population of decision makers (DMs) as if each DM maximizes her utility, which is an independent identically distributed draw from a distribution of preferences, over their choice set (i.e., menus). This theory is usually referred to as random utility model (RUM).\footnote{RUM was originally formulated by \citet{block_random_1960} and \citet{falmagne_representation_1978}.} If RUM is successful at describing behavior, then the analyst can use it to recover the distribution of heterogeneous preferences solely from observing the probability of choice of a finite set of alternatives from different menus justifying the name of revealed preference.\footnote{Since \citet{fishburn1998stochastic}, it is known that only partial identification of the distribution of preferences is possible under RUM. The probability of an item being ranked first is uniquely identified \citep{ABDsatisficing16}.} This distribution of preferences is an important input for many social sciences and can play a key role in policymaking. However, RUM may fail at describing behavior if DMs do not consider all available alternatives. This may happen, for instance, if there is a cost to understanding the decision task. In this situation, DMs may use a two-stage procedure:\footnote{Since \citet{manzini2007sequentially}, there has been a renewed interest in studying models of sequential choice in economics. The main aim of this research is to accommodate context effects that produce violations of the standard utility maximization framework. In a sequential procedure of choice or a two-stage choice procedure, the DMs first simplify the original choice problem using some heuristic and then choose rationality from the simplified choice problem. Limited consideration is just one example of a factor that affects the first stage, which determines the effective choice set used in the second stage. Other factors may be  willpower and status-quo bias \citep{horan2016simple}.} first selecting a subset of the given menu (consideration set), and only then choosing the best alternative from that set. Given that there may be latent heterogeneity in DMs preferences, and in how DMs form consideration sets, from the analyst's standpoint, both the consideration sets and choices from these sets are random. As a result of this two-stage procedure, if the consideration set does not contain the most preferred alternative of a DM, the DM will choose a dominated alternative, failing to be consistent with RUM.\footnote{For examples of the distortions created by limited consideration, see \citet{ho2017impact} and \citet{heiss2016inattention} (the health insurance market), \citet{hortaccsu2017power} (the residential electricity market in Texas), \citet{honka2014quantifying} and \citet{honka2017advertising} (the US auto insurance and banking industries), \citet{santos} (web browsing behavior of consumers when shopping online), \citet{molinari2018} (insurance purchases).}
\par
This paper proposes a generalization of the stochastic revealed preference methodology that is robust to limited consideration, allows for heterogeneous preferences that can be correlated with consideration, and is amenable to statistical testing. In doing so, we provide nonparametric restrictions on limited consideration and preferences that make partial recoverability of the distribution of preferences possible in a large class of stochastic choice data sets. Similar to \citet{kitamura2019nonparametric}, our framework also permits robust counterfactual welfare analysis and out-of-sample predictions.
\par
A large literature, pioneered by \citet{masatliogluRA} and \citet{manzini2014stochastic}, has proposed theories of consideration-mediated choice. These theories accommodate some departures from RUM caused by inattention, feasibility,
categorization, and search.\footnote{See, for instance, \citet{ABDsatisficing16, brady2016menu, caplin2016rationalconsideration, aguiar2017random, kovach2017satisficing, lleras2017more},
and \citet{horan2018}.} However, in contrast to our work, most existing theories of random consideration have assumed that preferences are \emph{homogeneous} \citep{cattaneo2017random}. This restriction implies that these models are not well suited to describe behavior at the population level. Our framework allows for homogeneous and heterogeneous preferences that may be correlated with consideration, making it suitable for both experimental and field data sets.
\par
The closest paper to our work is \citet{cattaneo2017random}. They provide a general framework, Random Attention Model (RAM), to test different models of stochastic consideration when preferences are homogeneous. Therefore, their work is applicable to individual stochastic choice data. \citet{cattaneo2017random} impose a set-monotonicity restriction on the probability of considering a set of alternatives given the menu. Namely, they assume that the probability of considering a given set cannot increase if the menu is getting larger. We study the implications of imposing this set-monotonicity constraint as well, but we allow for heterogeneous preferences. Hence, our model is applicable to both individual and population stochastic choice data.
\par
The second assumption we impose is \textit{preference stability}. This condition requires that the marginal distribution of preferences does not depend on the menu. The same stability assumption is used in \citet{mcfadden1990stochastic} and it means that the variation in menus does not change the preferences of DMs. The stability assumption is satisfied in many empirical and theoretical settings (see Sections~\ref{sec: stability} and~\ref{sec: examples} for further details). Together with set-monotonicity, it extends the RAM framework to settings with preference heterogeneity and the RUM framework to settings with limited consideration. Importantly, stability puts no restrictions on the dependence structure between menus and consideration, is consistent with statistical dependence between random consideration and random preferences, and together with set-monotonicity makes our framework testable. We show that none of these assumptions alone has empirical content, but together they restrict behavior meaningfully and allow for welfare and counterfactual analysis.
\par
Set-monotonicity and stability are compatible with the behavior of a mixture of DMs, where each DM's behavior is consistent with stochastic limited consideration governed by a RAM (i.e., requiring set-monotonicity). 
Under this interpretation, stability requires that the heterogeneity of preferences of the population of DMs is independent of the choice set, as in the McFadden and Richter's \citeyearpar{mcfadden1990stochastic} stochastic revealed preference framework. 
\par
We also demonstrate that set-monotocity and stability are satisfied by several important models of limited consideration with heterogeneous preferences, such as a variant of Manzini and Mariotti's \citeyearpar{manzini2007sequentially} model of alternative specific consideration with heterogeneous preferences, Brady and Rehbecks's \citeyearpar{brady2016menu} model of logit attention with random utility, Tversky's \citeyearpar{tversky1972elimination} model of elimination by aspects, search and satisficing with random search and random utility \citep{ABDsatisficing16}, and a version of rational inattention with Shannon's cost of information \cite{caplin2016rational}.\footnote{\citet{aguiar2018does} considers the same primitives but imposes full independence between random preferences and attention. In contrast to this paper, \citet{aguiar2018does} requires the presence of a default alternative and imposes parametric restrictions on the random attention.}
\par
Our approach differs from previous works that have used enhanced data sets to test for the presence of consideration. In particular, we only need a standard stochastic choice data set widely used in the discrete-choice literature.\footnote{For examples of enriched data sets that identify limited consideration, see \citet{reutskaja2011search} (eye-tracking data); \citet{honka2017advertising} and \citet{draganska2011choice} (additional surveys);  \citet{kawaguchi2016identifying} and \citet{conlon2013demand} (variation in product availability);  \citet{dehmamy2014utility} and \citet{huang2018pennies} (variations in quantity purchased and products purchased); and \citet{gabaix2006costly} (mouse-tracking data).} Recently, \citet{abaluck2017consumers} use structural restrictions on the elasticity behavior of demand to identify consideration sets and preferences. We differ from that work because we do not observe attributes (e.g., prices). \citet{molinari2018} obtain information about parametric distribution of preferences in a domain with attributes variation by introducing a support restriction on possible consideration sets. Our framework does not impose any parametric restrictions on the distribution of preferences, and allows both shape and support restrictions on consideration probabilities. \citet{kashaev2019peer} develop a dynamic model of discrete choice that incorporates peer effects into random consideration sets. They identify  preferences and consideration probabilities in a fixed menu settings by using variation in choices of peers. We assume menu variation and do not have access to panel data.
\par
\citet{aguiar2021} study nonparametric identification and estimation of the distribution of consideration sets and preferences without menu variation in panel data settings. 
\citet{dardanoni2020inferring} provide identification of the consideration probabilities given a known distribution of preferences in a fixed menu. They also consider grouped data sets where three instances of choice of the same consumers is observed to enhance identification. We assume menu variation, do not need to know the distribution of preferences, and do not use enriched stochastic choice data sets. More recently, \citet{dardanoni2020mixture} provide identification arguments for both preferences and cognition heterogeneity (including consideration probabilities) in mixture data sets. In contrast to our work, their method requires observing the joint distribution of choice over different menus. Also, their results are focused on parametric heterogeneity.
\par
The paper is organized as follows. Section~\ref{sec: model} introduces our general framework. In Section~\ref{sec: stability}, we provide several justifications of the stability assumption. Section~\ref{sec: examples} shows that our framework generalizes several important models of limited consideration. Section~\ref{sec: characterization} provides the characterization of our model which is  amenable for statistical testing, and discusses the computational aspects of our model. Section~\ref{sec: conditions} studies the implications of our model for preference revelation and counterfactual welfare analysis. Section~\ref{sec: conclusion} concludes our paper. All proofs can be found in Appendix~\ref{app: proofs}.

\section{Model}\label{sec: model}
Let $X$ be a finite choice set. The collection of choice sets (menus) is denoted by a nonempty subset of the power set $\mathcal{A}\subseteq2^{X}\setminus\{\emptyset\}$. We define the stochastic choice function $\rho_{A}\in\Delta(A)$, where $\Delta(A)$ denotes the set of all probability distributions on $A$, for $A\in\mathcal{A}$ such that $\rho_{A}(a)$ denotes the probability of choosing $a\in A$. The stochastic choice data set is the vector $\rho=(\rho_{A})_{A\in\mathcal{A}}$.
We call a stochastic choice data set complete if $\mathcal{A}=2^{X}\setminus\{\emptyset\}$ and incomplete otherwise. 
\par 
We let $U\subseteq X\times X$ be the set of linear orders (strict preference relations) defined on $X$. The typical element will be denoted by $\succ\in U$. 
\par
Within our framework, DMs may exhibit limited consideration. DMs exhibit limited consideration when they maximize their preferences in a strict subset of the menu. This strict subset is called a consideration set. We model limited consideration using the notion of consideration filters.

\begin{definition}[Consideration Filter]
We say that $\phi:2^X\setminus\{\emptyset\}\to2^{X}$ is a feasible (consideration) filter if there exists $D\in 2^{X}\setminus\{\emptyset\}$ such that
    \[
    \phi(A)=\begin{cases}
    D, &D\subseteq A,\\
    \emptyset, &\text{otherwise}
    \end{cases}
    \]
    for all $A\in2^X\setminus\{\emptyset\}$.
\end{definition}
Let $\Phi$ be a finite collection of all feasible filters. The typical element of it will be denoted by $\phi\in \Phi$.
\par
We consider a random attention and utility model (RAUM). A \emph{behavioral type} of this model is determined by a pair of preferences and filter $(\succ,\phi)\in U\times \Phi$. A RAUM rule $\pi=(\pi_{A})_{A\in2^X\setminus\{\emptyset\}}$ is a collection of probability distribution over preferences and filters $\pi_{A}\in\Delta(U\times \Phi)$ such that $\pi_{A}(\succ,\phi)=0$ whenever $\phi(A)=\emptyset$ for all choice sets $A\in2^X\setminus\{\emptyset\}$.
\par
Given menu $A$, let $\pi_A(\cdot|\succ)$ denote the conditional distribution over filters conditional on the random preference order being $\succ$. Then, for $\phi$ such that $\phi(A)=D$, $\pi_A(\phi|\succ)$ is the probability that set $D$ is considered in menu $A$ by DMs with preferences $\succ$. Essentially, consideration filters are indexed by all possible consideration sets $D$: different filters will generate all subsets of a given menu $A$. 
\par
We work with filters that are indexed by consideration sets because of two main reasons. First, we think of $\rho$ as coming from repeated cross-sections. (See the experiment in \citealp{aguiar2018does} for an example of a setting where each DM faces a menu at random and has only one choice instance.) Thus, for a given menu $A$, DM with filter $\phi$ considers $\phi(A)$ and a pair $(\succ,\phi)$ completely describes her behavior. For a different menu, a different DM may be endowed with a different filter and preferences. Second, consideration filters are convenient since most of the models of limited consideration and assumptions about them are defined in terms of probabilities of considering a given set $D$. This greatly simplifies our notation and mathematical exposition without loss of generality. 
\par 
Note that our RAUM rule is empirically equivalent to another rule that uses alternative behavioral types consisting of a preference and a consideration mapping that maps menus to their nonempty subsets. An example of such a consideration mapping is an attention filter \citep{masatliogluRA}. Our RAUM rule is be a mixture of these behavioral types. This alternative representation may be more natural for some readers but ultimately it is fully exchangeable with ours (see Section~\ref{sec: stability}). Moreover, our representation is more mathematically convenient because it allows us to state the main restrictions on a RAUM rule in a way that is substantially more efficient and friendlier for computational implementation.\footnote{For example, our representation is more convenient in imposing restrictions on a conditional probability of considering a set conditional on a preference order and a menu. In particular, our representation does not require summations over all consideration mappings that map a given menu to a consideration set.} 
\begin{definition} A stochastic choice data set $\rho$ admits a RAUM representation $\pi$ if
\[
\rho_{A}(a)=\sum_{(\succ,\phi)\in U\times \Phi}\pi_{A}(\succ,\phi)\Char{a\in\phi(A),\:a\succ b,\:\forall b\in\phi(A)\setminus\{a\}},
\]
for all $a\in A$ and all $A\in\mathcal{A}$.
\end{definition} 
RAUM is so general that it does not have any empirical content. That is, feasible filters together with unrestricted (possibly menu dependent) distribution of preferences are permissive enough to explain any behavior. Hence, without further constraints, it is impossible to falsify RAUM or to recover the (marginal) distribution of preferences of a population of DMs (i.e., $\pi^*_{A}(\succ)=\sum_{\phi}\pi_{A}(\succ,\phi)$). We impose the following stability constraint on the RAUM representation $\pi$.

\begin{assumption} [Stability] There exists $\pi^{*}\in\Delta(U)$ such that $\pi^*_{A}(\succ)=\pi^{*}(\succ)$ for any $A\in2^X\setminus\{\emptyset\}$ and $\succ\in U$. 
\end{assumption}

Note that stability is equivalent to requiring that $\pi^*_{A}(\succ)=\pi^*_{B}(\succ)$ for any $A,B\in2^X\setminus\{\emptyset\}$ and $\succ\in U$, thus, justifying its name. 
\par
One interpretation of stability is that it restricts limited consideration such that the marginal distribution of preferences of the general RAUM is equivalent to the \emph{true} distribution of heterogeneous preferences in the population. The true distribution of preferences is the distribution on $U$ that controls behavior in the counterfactual situation of absence of limited consideration. Stability does not require the knowledge of such distribution. Moreover, our stability assumption is a natural analogue, within our more general framework, of the assumption of preference stability in the stochastic rationality model of \citet{mcfadden1990stochastic}.
\par 
Importantly, stability is consistent with stochastic dependence between consideration filters and random preferences. We only require that the (marginal) distribution of preferences remains the same across exogenously given menus of alternatives. We further explore limitations of the stability assumption in the next section (Section~\ref{sec: stability}).
\par
Even under stability, limited consideration has to be further restricted to have empirical bite as we will show in Proposition~\ref{prop: emp bite of st and mon}. Here, we follow \citet{cattaneo2017random} and impose the following restriction.

\begin{assumption}[Set-monotonicity] 
For any $\succ$, $\phi$, $A$, and $B$ such that $A\subseteq B$ and $\phi(A)\neq\emptyset$, it must be that $\pi_{A}(\phi|\succ)\geq\pi_{B}(\phi|\succ)$.
\end{assumption} 

Set-monotonicity means that the conditional probability of a given filter, $\phi$, conditional on a preference type, $\succ$, cannot increase as the menu expands. That is, DMs will pay more attention to a set when the menu of alternatives is smaller. Intuitively, larger menus have a higher opportunity cost of consideration. \citet{cattaneo2017random} show that many models of random consideration satisfy set-monotonicity. 
\par
The interaction of set-monotonicity and stability does not imply independence of consideration and preferences, as the following example demonstrates. 
\begin{example}\label{example: dependence}
Let $X=\{a,b\}$ and $\mathcal{A}=\{\{a\},\{b\},\{a,b\}\}$. Let $a\succ_1 b$ and $b\succ_2 a$, and assume that only two filters below realize with nonzero probability: 
\[
    \phi_1(A)=\left\{\begin{array}{cc} 
    \{a\}, & \{a\}\subseteq A,\\
    \emptyset, & \text{otherwise},
    \end{array}\right.\qquad
    \phi_2(A)=\left\{\begin{array}{cc} 
    \{b\}, & \{b\}\subseteq A,\\
    \emptyset, & \text{otherwise}.
    \end{array}\right.\qquad
\]

Consider the following distributions $\pi_{A}(\succ,\phi)$ over the above two preference orders and two filters for different menus:
\begin{equation*}
\begin{tabular}{c|c|c|c|c|}
\multicolumn{1}{c}{{$\pi_{A}(\succ,\phi)$}}&\multicolumn{1}{c}{$(\succ_1,\phi_1)$}&\multicolumn{1}{c}{$(\succ_1,\phi_2)$}&\multicolumn{1}{c}{$(\succ_2,\phi_1)$}&\multicolumn{1}{c}{$(\succ_2,\phi_2)$}\\
\cline{2-5}
$\{a,b\}$ & $1/3$ & $1/6$& $1/6$ & $1/3$\\  
\cline{2-5}
$\{a\}$ & $1/2$ & $0$& $1/2$ & $0$\\ 
\cline{2-5}
$\{b\}$ & $0$ & $1/2$& $0$ & $1/2$\\ 
\cline{2-5}
\end{tabular}
\end{equation*}
Note that $\pi^*_{A}(\succ_1)=\pi^*_{A}(\succ_2)=1/2$ for all $A$. However,
\[
\dfrac{1}{3}=\pi_{\{a,b\}}(\succ_1,\phi_1)\neq\pi^*_{\{a,b\}}(\succ_1)\sum_{i=1}^2\pi_{\{a,b\}}(\succ_i,\phi_1)=\dfrac{1}{2}\cdot\dfrac{1}{2}=\dfrac{1}{4}.
\] 
That is, preferences and filters are not independent. Moreover, set-monotonicity is also satisfied. For instance, $\pi_{\{a\}}(\phi_1|\succ_1)=1\geq2/3=\pi_{\{a,b\}}(\phi_1|\succ_1)$.
\end{example}
\par
The next proposition qualifies the importance of stability and set-monotonicity working together. Neither of these restrictions alone are enough for empirical relevance of the model. However, when they are combined together, the model becomes falsifiable even with limited menu variation. We see the combination of these two restrictions as a baseline of empirical content that makes our study empirically meaningful. 

\begin{proposition}\label{prop: emp bite of st and mon}
The following statements are true: 
\begin{enumerate}
\item Any $\rho$ admits a stable RAUM representation. 
\item Any $\rho$ admits a set-monotone RAUM representation.
\item There exists an incomplete $\rho$ that does not admit a set-monotone and stable RAUM representation.
\end{enumerate}
\end{proposition} 

Here we provide a sketch of the proof of (iii). We construct an incomplete data set (i.e., $\mathcal{A}\neq 2^X\setminus\{\emptyset\}$) that does not admit a set-monotone and stable RAUM. Let $X=\{a,b,c,d\}$ and
\[
\mathcal{A}=\{\{a,b\}, \{a,c\},\{b,d\}, \{a,b,d\},\{a,c,d\}, \{b,c,d\}\}.
\]
Suppose the observed $\rho$ is as follows
\begin{equation*}
\begin{tabular}{c|c|c|c|c|c|c|}
\multicolumn{1}{c}{}&\multicolumn{1}{c}{$\{a,b\}$}&\multicolumn{1}{c}{$\{a,c\}$}&\multicolumn{1}{c}{$\{b,d\}$}&\multicolumn{1}{c}{$\{a,b,d\}$}&\multicolumn{1}{c}{$\{a,c,d\}$}&\multicolumn{1}{c}{$\{b,c,d\}$}\\
\cline{2-7}
$\rho_A(a)$ & $1$ & $1$ & - & $0$ & $0$& -\\ 
\cline{2-7}
$\rho_A(b)$ & $0$ & - & $1$ & $1$ & - &$\alpha_b$\\ 
\cline{2-7}
$\rho_A(c)$ & - & $0$ & - & - & $1$& $\alpha_c$\\ 
\cline{2-7}
$\rho_A(d)$ & - & - & $0$ & $0$ & $0$& $\alpha_d$\\ 
\cline{2-7}
\end{tabular}
\end{equation*}
where $\alpha_d>0$. Consider the pair $\{a,b\}$ and $\{a,b,d\}$. From observing $\rho_{\{a,b\}}(b)=0$ and $\rho_{\{a,b,d\}}(b)=1$, we can conclude that $b\succ d$ with probability $1$ or $b$ and $d$ are never considered together. Similarly, from observing $\rho_{\{a,c\}}(c)=0$ and $\rho_{\{a,c,d\}}(c)=1$, we can make analogous conclusion about $c$ and $d$. As a result, the fact that $\alpha_d>0$ then implies that the probability of considering the singleton consideration set $\{d\}$ must be nonzero (otherwise $d$ is either never considered or dominated by $b$ or $c$). But the latter is impossible because in menu $\{b,d\}$ option $d$ is never chosen. The formal details of the sketch above can be found in Appendix~\ref{app: emp bite of st and mon}.

\section{Stability of Preferences as Structured Heterogeneity in a Population of Inattentive DMs}\label{sec: stability}
The set-monotonicity property is well-understood and justified due to \citet{cattaneo2017random}. Here, we provide several justifications of stability as a reasonable assumption for the RAUM representation.
\par
One possible interpretation of the set-monotone and stable RAUM representation is that it represents a mixture of the behavior of a population of DMs, where the behavior of each DM is consistent with RAM.

\begin{definition} [Random Attention Model, RAM, \citealp{cattaneo2017random}]
A stochastic choice data set $\rho$ admits a RAM representation if there exist a preference order $\succ$ and a collection of distributions over consideration filters (an attention rule) $\lambda_\succ=(\lambda_{A,\succ})_{A\in\mathcal{A}}\in \Delta(\Phi)^{\abs{\mathcal{A}}}$, where $\abs{\mathcal{A}}$ is the cardinality of $\mathcal{A}$, such that $\lambda_{A,\succ}(\phi)=0$ whenever $\phi(A)=\emptyset$; $\lambda_{A,\succ}(\phi)\geq\lambda_{B,\succ}(\phi)$ for all $A,B\in\mathcal{A}$ such that $A\subseteq B$ and $\phi(A)\neq\emptyset$; and 
\[
\rho_{A}(a)=\sum_{\phi}\lambda_{A,\succ}(\phi)\Char{a\in\phi(A),\:a\succ b,\:\forall b\in\phi(A)\setminus\{a\}}
\]
for each $A\in\mathcal{A}$.
\end{definition}

To better understand the relation between RAM and RAUM, consider the following data generating process for $\rho$. Fix some distribution over preference orders $\pi^*\in\Delta(U)$ and some collection of attention rules for all possible preference orders $\{\lambda_{\succ}\}_{\succ\in U}$. Every DM draws a preference order $\succ$ from $\pi^*$ independently of other DMs, as in the \citet{mcfadden1990stochastic}'s framework. Given the preference order $\succ$ and attention rule $\lambda_\succ$, the DM chooses alternatives from $A$ according to the RAM rule (that induces a probability of choice) 
\[
\rho_{A,\succ}(a)= \sum_{\phi}\lambda_{A,\succ}(\phi)\Char{a\in\phi(A),\:a\succ b,\:\forall b\in\phi(A)\setminus\{a\}}.
\]
Let $\rho$ be a mixture of the above RAM rules weighted by $\pi^*$. That is,
\[
\rho_{A}(a)=\sum_{\succ} \pi^*(\succ)\rho_{A,\succ}(a)
\]
for each $A\in\mathcal{A}$. Note that this data set admits a set-monotone and stable RAUM representation $\pi$ with $\pi_A(\succ,\phi)=\lambda_{A,\succ}(\phi)\pi^*(\succ)$ for all $\succ$, $\lambda$, and $A$. Moreover, by construction, any $\rho$ that admits a set-monotone and stable RAUM $\pi$ is also a mixture of RAM rules induced by $\pi^*$ and attention rules $\{\pi_A(\cdot|\succ)\}_{\succ\in U}$. In other words, the set-monotone and stable RAUM framework is an extension of the RAM framework to heterogeneous preferences structured as in the \citet{mcfadden1990stochastic}'s framework.
\par
Alternatively, since $\rho$ admits a RUM representation if there exists a distribution over preference orders $\pi^*\in\Delta(U)$ such that
\[
\rho_{A}(a)=\sum_{\succ}\pi^*(\succ)\Char{a\succ b,\:\forall b\in A\setminus\{a\}},
\]
we can think of RUM rules as mixtures of RAM rules obtained from particular degenerate attention rules (i.e., $\lambda_{A,\succ}(\phi)=\Char{\phi(A)=A}$). Hence, the set-monotone and stable RAUM framework is an extension of the RUM framework to heterogeneous consideration filters, as in the \citet{cattaneo2017random}'s framework.\footnote{Note that this interpretation of the stable and set-monotone RAUM rule means that we could alternatively and equivalently define our behavioral types as a pair of a preference and a consideration mapping, describing the nonempty consideration sets for each menu. Indeed, following \citet{cattaneo2017random} we can restrict the consideration mapping to be an attention filter. The attention filter property states that if we remove an item from a menu that is not in the deterministic consideration set, then the consideration set in the new menu is the same as the consideration set in the original menu.}
\par
It is important to point out that in this generalization of the RAM framework, which allows for heterogeneous preferences, we use the same stability assumption as in the classical stochastic revealed preference framework of \citet{mcfadden1990stochastic}. \citet{kitamura2018nonparametric} have given a modern interpretation to this assumption as an exogeneity restriction requiring preferences and menus to be independent. There are many settings where this assumption is satisfied:  (i) experiments, where the experimenter exogenously varies the menus \citep{aguiar2018does}; (ii) environments with predetermined choice sets in time (e.g., such as modes of transport or elections \citealp{mcfadden1986choice}); (iii) choice problems with frequently-purchased and inexpensive products \citep{lu2021estimating}.\footnote{Note that if there are observable covariates (e.g., product characteristics), then our analysis goes through after conditioning on such covariates. In that case, stability would require preferences to be independent of menus conditional on observed covariates.} Moreover, the stability assumption has been the standard in decision theory with menu variation since the work of \citet{falmagne_representation_1978}. 
\par
An important reason why the distribution of preferences is assumed to be independent of menus is that this restriction allows for well-defined and informative welfare and counterfactual choice analysis out-of-sample (i.e., in menus that are not part of the data set). This stable distribution over preferences or types can be interpreted as a \textbf{true preference distribution}. This follows the tradition of the seminal consideration set papers in assuming that each DM is endowed with a menu independent strict preference relation \citep{masatliogluRA,manzini2014stochastic,cattaneo2017random}. 
\par 
The stability assumption may not be suitable for all choice situations. A classical example arises in survey data, such as the application studied in \citet{kitamura2018nonparametric}. Stability may fail because income, which determines the budget faced by a DM, may be correlated with preferences. In these situations, \citet{kitamura2018nonparametric} suggest using the control function approach \citep{blundell2001endogeneity,imbens2009identification} to suitably modify the stochastic revealed preference framework in \citet{mcfadden1990stochastic}. A study of the connection between their solution to the problem of endogeneity and our framework is left for future work.

\section{Examples of RAUM}\label{sec: examples}
Notwithstanding the apparent restrictiveness of the stability and set-monotonicity assumptions, RAUM generalizes a wide variety of models of limited consideration and random utility, allowing for correlation between attention and preferences, as showcased in the next examples.  

\subsection*{Attention-index Models}
Given $\succ$, an attention-index $\eta_{\succ}:2^{X}\to\Real_{+}$ such that $\eta_{\succ}(\emptyset)=0$ is a capacity  over subsets of $X$ capturing how enticing they are conditional on a given preference type. The following models of consideration are examples of rules that are governed by attention-indexes. These type of models are studied in \cite{aguiar2018does} with an independence assumption between preferences and consideration. \cite{abaluck2017consumers} also uses these type of models in a different domain with the same independence assumption. Here, we show that models with an attention-index consideration that depends on heterogeneous preferences admit a set-monotone and stable RAUM representation. 
\begin{example}[Attention-index\label{ex: Attentionindex}]
Assume that there is a vector of random utilities associated with $X$, $\rand{u}=(\rand{u}_y)_{y\in X}$ and a vector of random saliency $\rands{\xi}=(\rands{\xi}_y)_{y\in X}$ with a menu-independent cumulative distribution function (c.d.f.) $F_{(\rands{\xi},\rand{u})}$. Moreover, assume that the marginal c.d.f. of $\rand{u}$, $F_{\rand{u}}$, is continuous.\footnote{Continuity of $F_{\rand{u}}$ implies that $\rand{u}_{y}=\rand{u}_{y'}$ with probability zero for any $y\neq y'$.} 
Let $\mathcal{U}_{\succ}=\{u\in \Real^{\abs{X}}\::\: \forall y,y'\in X,u_{y}>u_{y'}\iff y\succ y'\}$ be the set of utility values that imply $\succ$ \citep{fishburn1998stochastic}. Then, for every $\succ\in U$, define an attention-index $\eta_{\succ}$ as
\[
\eta_{\succ}(D)=\int\Char{u\in \mathcal{U}_{\succ}}\abs{\{y\in D\::\:u_y+\xi_{y}\geq\kappa\}}dF_{(\rands{\xi},\rand{u})}(\xi,u)
\]
for some $\kappa\in\Real$.\footnote{The support of the random variables and $\kappa$ have to be such that  $\eta_{\succ}(D)>0$ for all $D$ and $\succ$.} The attention index $\eta_{\succ}$ captures the attractiveness of $D$ by computing the average cardinality of the set of all alternatives in $D$ that are above a threshold $\kappa$.\footnote{When $\kappa\rightarrow -\infty$ the attention-index converges to $|D|$ and there is no dependence on preferences. This special case has been explored in \citet{aguiar2018does} in the presence of a default alternative.} Next, we let $\pi_A^*$ be defined as 
\[
\pi^*_A(\succ)=\int \Char{u\in \mathcal{U}_{\succ}}dF_{\rand{u}}(u)
\]
for all $\succ\in U$. 
Finally, for any $\phi$ and $A$ such that $\phi(A)\neq\emptyset$, we define $\pi_A(\phi|\succ)$ as 
\[
\pi_A(\phi|\succ)=\dfrac{\sum_{B\subseteq X\setminus \tilde{g}(A)}\eta_{\succ}(\phi(A)\cup B)}{\sum_{C\in 2^A\setminus\{\emptyset\}}\sum_{B\subseteq X\setminus \tilde{g}(A)}\eta_{\succ}(C\cup B)}
\]
for a known mapping $\tilde{g}:2^X\to 2^X$.
\end{example}
In Example~\ref{ex: Attentionindex}, stability holds because the marginal c.d.f. $F_{\rand{u}}$ does not depend on the menu. In the absence of limited consideration, the distribution over preferences $\pi^*$ constitutes a random utility rule, as in \citet{mcfadden1990stochastic}. For instance, if $\abs{X}=3$, $\rand{u}_y=w(y)+\rands{\epsilon}_y$ for some mean-utility function $w:X\to \Real$, and $\epsilon$ is distributed according to the Gumbel distribution, then the random utility (marginal) takes the logit form and
\[
\pi^*_{A}(\succ)=\frac{e^{w(y_1)}}{\sum_{y\in X}e^{w(y)}}\frac{e^{w(y_2)}}{e^{w(y_2)}+e^{w(y_3)}},
\]
for all $y_1,y_2,y_3\in X$ and $\succ\in U$ such that $y_1\succ y_2\succ y_3$.
\par 
Set-monotonicity holds for two important special cases in the literature: (i) the logit attention model \citep{brady2016menu} with $\tilde{g}(A)=X$ for all $A$; (ii) and the elimination by aspects \citep{tversky1972elimination,aguiar2017random} with $\tilde{g}(A)=A$ for all $A$.\footnote{See Appendix~\ref{app: gen formula} for additional details about these special cases. Sometimes we have an alternative that is present in every menu--the default alternative. The eliminations by aspects model is also called Random Categorization Rule \citep{aguiar2017random} when there is a default alternative that always attracts attention.}  The item-specific attention model \citep{manzini2014stochastic} is also another special case \citep{suleymanov}.\footnote{In Appendix~\ref{app: gen formula}, we provide a restriction on $\tilde{g}$ that guarantees set-monotonicity. This condition is satisfied by all these models.} 
\par
We highlight that correlation between random saliency and random utility is not restricted (i.e., $F_{(\rands{\xi},\rand{u})}$ is unrestricted). Finally, in Example~\ref{ex: Attentionindex}, we use the cardinality of the set of salient alternatives
\[
\abs{\{y\in D\::\:u_y+\xi_{y}\geq\kappa\}}
\]
to construct $\eta_{\succ}(D)$.
However, any other mappings can be used to construct $\eta_{\succ}$ without the resulting model deviating from the stable and set-monotone RAUM framework. 

\subsection*{Search and Satisficing}
The search and satisficing behavioral model \citep{simon1955behavioral} is an important example of a model with dependence between preferences and consideration driven by random utility directly affecting the way choice sets are formed.

\begin{example}[Search and Satisficing]\label{ex: SS}
Given a choice set $X$, let $\rand{s}=(\rand{s}_y)_{y\in X}$ be a random search index and $\rand{u}=(\rand{u}_y)_{y\in X}$ be a random utility with a joint continuous c.d.f. $F_{(\rand{s},\rand{u})}$. Similar to Example~\ref{ex: Attentionindex}, the random vector $\rand{u}$ induces $\pi^*$ such that
\[
\pi^*_A(\succ)=\int \Char{u\in \mathcal{U}_{\succ}}dF_{\rand{u}}(u)
\]
for all $\succ\in U$. An alternative $y_1$ is searched earlier than $y_2$ if and only if its search index is bigger (i.e., $s_{y_1}\geq s_{y_2}$). All DMs face a common threshold $\tau\in \Real$, and their preferences are captured by random utility $\rand{u}$. For every $\succ\in U$ define $\pi_{A}(\cdot|\succ)$ as the conditional distribution over filters such that $\pi_{A}(\phi|\succ)=0$ if $\phi(A)=\emptyset$ and for $\phi(A)\neq\emptyset$
\[
\pi_{A}(\phi|\succ)=\int \Char{u\in\mathcal{U}_{\succ},\phi(A)=\{y\in A\:|\: \not{\exists} y'\in A : s_{y}<s_{y'}, u_{y'}\geq\tau\}}dF_{(\rand{s},\rand{u})}(s,u),
\]
where $\tau\in\Real$ is a threshold that is common for all DMs. We assume $F_{(\rand{s},\rand{u})}$ is continuous to avoid indifference in preferences and to avoid the case that many items are searched simultaneously. In words, the probability of the consideration set being equal to $\phi(A)=D$ is given by the probability of items in $D$ being searched before the items whose utility value are above the threshold in a menu. If no item is satisficing, then the whole menu $A$ is searched.  
\end{example}

The rule $\pi(\succ,\phi)=\pi_{A}(\phi|\succ)\pi_A^*(\succ)$ from Example~\ref{ex: SS} is set-monotone because the threshold is constant and the distribution of the search index does not depend on the menu.\footnote{This example fits Example $9$ in the supplement of \citet{cattaneo2017random}. But it is easy to see that a bigger menu (i.e., $B\supseteq A$) will only decrease the probability of filter $\phi$ (i.e., $\pi_B(\phi|\succ)\leq \pi_A(\phi|\succ)$).} In addition, stability is satisfied because the joint distribution of the random utility and the search index does not depend on the menu. This version of search and satisficing was studied and characterized by \citet{ABDsatisficing16}. The interpretation of this particular RAUM is compatible with $\rho$ being generated by the behavior of a population of DMs. Each of these DMs chooses according to random satisficing behavior (with a fixed distribution of search) as in \citet{ABDsatisficing16}. The preference heterogeneity is governed by $\pi^*$. 
\par
Correlation between preferences and consideration arises endogenously due to the use of the utility to stop the search process once a satisficing item is found.  \citet{caplin2011search} showed that search and satisficing behavior, as in our example, can be written as an optimization problem that produces optimal consideration sets given search costs and expected value of each alternative. In that sense, correlation between preferences and consideration is the result of optimizing behavior of DMs.   
\par
Set-monotonicity breaks if we allow menu-dependent thresholds \citep{aguiarkimya2019adaptive} or menu-dependent search indexes \citep{ABDsatisficing16}. However, if the threshold is random\footnote{$\tau$ becomes random and there is a joint c.d.f. $F_{(\rands{\tau},\rand{s},\rand{u})}$ governing the model.}, as in \citet{kovach2017satisficing}, then set-monotonicity is satisfied if the distribution of the random threshold is menu independent. 

\subsection*{Rational Inattention} 
\citet{caplin2016rationalconsideration} pointed out the relationship between the theory of rational inattention in discrete choice and the theory of consideration sets. Even though the domain of the theory of rational inattention as presented in \citet{caplin2016rationalconsideration} is described in a different domain than the one over which our theory is developed, we show next that some special cases of the rational inattention model admit a set-monotone and stable RAUM representation. The first example of this section follows the consumer problem $1$ in \citet{caplin2016rationalconsideration}, but in contrast to them, (i) we allow for menu variation, and (ii) we assume that priors change with the menu in a way consistent with Bayes' rule. 

\begin{example}[Rational Inattention]\label{ex: RI}
Let $X=\{a,b,c\}$ be the choice set and the (unobserved) state space be equal to the choice set $\Omega=X$. Consider an individual DM. Let $\mu(y)$, $y\in X$, be the prior that $y$ is of high quality and the rest of alternatives are of low quality. Without loss of generality, assume that $\mu(a)\geq \mu(b)\geq \mu(c)$. Assume that, given menu $A\subseteq X$, the prior belief about the state of the world is updated according to Bayes' rule. That is, the prior that $y$ in menu $A$ is of the high quality is $\mu_A(y)=\mu(y)/\sum_{y'\in A}\mu(y')$. 
\par
There is a state dependent utility $u:X\times \Omega\to\{0,1\}$ such that, $u(a,y)=u_G$ if $a=y$ and otherwise $u(a,y)=u_B$ with $u_G>u_B$. The DM expends attention effort following Shannon's model of rational inattention with a fixed parameter $\lambda>0$.\footnote{We consider several values of $\lambda$ as a comparative statics analysis, but we assume that the DM has a fixed cost of attention effort.}
For tractability, define a net payoff of identifying the high-quality item that takes into account the attention effort as
\[
\delta=\exp\left(\frac{u_{G}-u_{B}}{\lambda}\right)-1.
\]
Define also,
\begin{align*}
    \delta_1^*=-3+\dfrac{1}{\mu(c)},\quad \delta_2^*=-1+\dfrac{\mu(a)}{\mu(c)},\quad \delta_3^*=-1+\dfrac{\mu(b)}{\mu(c)},\quad \delta_4^*=-1+\dfrac{\mu(a)}{\mu(b)}.
\end{align*}
Note that since $1-2\mu(c)=\mu(a)+\mu(b)-\mu(c)\geq\mu(a)$ and $\mu(a)\geq\mu(b)\geq\mu(c)$, it follows that $\delta^*_1\geq\delta^*_2\geq\delta^*_3$ and $\delta^*_2\geq\delta^*_4$. 
\citet{caplin2016rationalconsideration} show that rational inattention produces a deterministic consideration set that remains the same across states (i.e., for every $A$, there exists $\phi^*$ such that $\pi_A(\phi^*|\succ)=1$ for all $\succ\in U$). Tables~\ref{tab:RI1} and~\ref{tab:RI2} display the deterministic considerations set (i.e., $\phi^*(A)$ with probability $1$) for different values of $\delta$ and menu $A$. For instance, the $2\times 1$ element of Table~\ref{tab:RI1} is $\{a\}$, which corresponds to $A=\{a,b\}$ and $\delta\in (0,\delta^*_4]$. This means that if $A=\{a,b\}$ and $\delta\in (0,\delta^*_4]$, then the whole probability mass goes to the filter $\phi^*$ that is such that $\phi^*(\{a,b\})=\{a\}$. Table~\ref{tab:RI1} corresponds to the case when $\delta_3^*\geq \delta_4^*$ (i.e., $\mu(b)/\mu(c)\geq \mu(a)/\mu(b)$).
\begin{table}[h]
\centering
\begin{tabular}{c|c|c|c|c|c|}
\multicolumn{1}{c}{$A/\delta\in$}&\multicolumn{1}{c}{$(0,\delta^*_4]$}&\multicolumn{1}{c}{$(\delta^*_4,\delta^*_3]$}&\multicolumn{1}{c}{$(\delta^*_3,\delta^*_2]$}&\multicolumn{1}{c}{$(\delta^*_2,\delta^*_1]$}&\multicolumn{1}{c}{$(\delta^*_1,+\infty)$}\\
\cline{2-6}
$\{a,b,c\}$ & $\{a\}$ & $\{a,b\}$ & $\{a,b\}$ & $\{a,b\}$ & $\{a,b,c\}$\\ 
\cline{2-6}
$\{a,b\}$ & $\{a\}$ & $\{a,b\}$ & $\{a,b\}$ & $\{a,b\}$ & $\{a,b\}$\\ 
\cline{2-6}
$\{b,c\}$ & $\{b\}$ & $\{b\}$ & $\{b,c\}$ & $\{b,c\}$ & $\{b,c\}$\\ 
\cline{2-6}
$\{a,c\}$ & $\{a\}$ & $\{a\}$ & $\{a\}$ & $\{a,c\}$ & $\{a,c\}$\\ 
\cline{2-6}
\end{tabular}
\caption{Consideration sets for different menus $A$ and $\delta$ when $\delta_3^*\geq \delta_4^*$.}
    \label{tab:RI1}   
\end{table}
Table~\ref{tab:RI2} considers the case when $\delta_4^*\geq \delta_3^*$.   
\begin{table}[h]
\centering
\begin{tabular}{c|c|c|c|c|c|}
\multicolumn{1}{c}{$A/\delta\in$}&\multicolumn{1}{c}{$(0,\delta^*_3]$}&\multicolumn{1}{c}{$(\delta^*_3,\delta^*_4]$}&\multicolumn{1}{c}{$(\delta^*_4,\delta^*_2]$}&\multicolumn{1}{c}{$(\delta^*_2,\delta^*_1]$}&\multicolumn{1}{c}{$(\delta^*_1,+\infty)$}\\
\cline{2-6}
$\{a,b,c\}$ & $\{a\}$ & $\{a\}$ & $\{a,b\}$ & $\{a,b\}$ & $\{a,b,c\}$\\ 
\cline{2-6}
$\{a,b\}$ & $\{a\}$ & $\{a\}$ & $\{a,b\}$ & $\{a,b\}$ & $\{a,b\}$\\ 
\cline{2-6}
$\{b,c\}$ & $\{b\}$ & $\{b,c\}$ & $\{b,c\}$ & $\{b,c\}$ & $\{b,c\}$\\ 
\cline{2-6}
$\{a,c\}$ & $\{a\}$ & $\{a\}$ & $\{a\}$ & $\{a,c\}$ & $\{a,c\}$\\ 
\cline{2-6}
\end{tabular}
\caption{Consideration sets for different menus $A$ and $\delta$ when $\delta_4^*\geq \delta_3^*$}
    \label{tab:RI2}   
\end{table}
\par
Tables~\ref{tab:RI1}-\ref{tab:RI2} imply that set-monotonicity is satisfied. Indeed, the deterministic consideration set described in each column of these tables satisfy the attention filter property described in \citet{masatliogluRA}.
\par
Following \citet{caplin2016rationalconsideration}, it can be shown that given a consideration set $D$, the distribution over choices for this rationally inattentive DM in $D$ is such that the probability of picking $y$ for a consideration set $D$ is
\[
P_{D}(y)=\dfrac{\mu_{D}(y)(|D|+\delta)-1}{\delta}.
\]
For instance, if $\delta\in(\delta_2^*,\delta_1^*]$ and $A=\{a,b,c\}$ (i.e, $D=\{a,b\}$), then $\rho_{\{a,b,c\}}(a)=P_{\{a,b\}}(a)$, $\rho_{\{a,b,c\}}(b)=P_{\{a,b\}}(b)$, and $\rho_{\{a,b,c\}}(c)=0$. Stability also holds. Since $\rho_A$ only depends on $A$ via consideration set $D$, for $\delta\leq\delta_1^*$, it is trivial to construct a menu independent distribution over preference orders over $X$. Hence,  $\rho$ generated by this rational inattention model admits a set-monotone and stable RAUM representation. 
\par
When $\delta>\delta_1^*$, DMs consider all alternatives. Hence, in this case, since there are only 3 alternatives, $\rho$ admits a RAUM representation if and only if $\rho$ is regular \citep{block_random_1960,falmagne_representation_1978}. That is, $\rho_A(y)\geq\rho_X(y)$ for all $A\subseteq X$ and $y\in A$. Note that $\rho_{\{a,b\}}(a)=\dfrac{\mu_{\{a,b\}}(a)(2+\delta)-1}{\delta}$ and $\rho_{\{a,b,c\}}(a)=\dfrac{\mu(a)(3+\delta)-1}{\delta}$. Hence,
\begin{align*}
    \rho_{\{a,b\}}(a)-\rho_{\{a,b,c\}}(a)=\dfrac{\mu(a)\mu(c)(\delta-\delta^*_1)}{\delta(\mu(a)+\mu(b))}>0,
\end{align*}
where the last inequality is implied by $\delta>\delta_1^*$. Similar inequalities can be derived for all other menus and options. For the formal construction of $\pi^*(\succ)$ see Appendix~\ref{app: RI1}.
\end{example}
\citet{caplin2016rationalconsideration} remark informally that the consideration sets produced by the Shannon's model of rational inattention satisfy the attention filter property first proposed in \citet{masatliogluRA}.
Deterministic consideration sets that are attention filters satisfy set-monotonicity \citep{cattaneo2017random}. However, this statement is only true under certain assumptions about how the prior is defined and how it changes when menus change. In Example~\ref{ex: RI}, the deterministic consideration set is indeed an attention filter. However, if we replace $\mu$ with another prior that arbitrarily depends on the menu, this property will not hold in general. 
\par 
Stability also holds in Example~\ref{ex: RI}. Note that randomness in choice for the rational inattentive DM is driven by mistakes due to costly attentional effort. If $\lambda=0$, then choice will be deterministic and consistent with utility maximization. The fact that stability holds in this setup is entirely due to the fact that Shannon's model of rational inattention admits an additive random utility equivalent representation, as stated in \citet{fosgerau2017discrete}. Similarly, this equivalence will hold only under certain assumptions on the priors and how they change across menus. If priors are fixed, this equivalence holds. In addition, we showed in our example another way to change priors that also makes this equivalence hold. In other words, stability holds here because we restrict the dependence of the priors on the menu, hence making it possible for the implied distribution over preferences $\pi^*$ to be menu-independent as stability requires. 
\par
Not all probabilities of choice generated by Shannon's rational inattention model admit a set-monotone and stable RAUM representation. A well-known example in \citet{matvejka2015rational} (see problem $4$) exhibits Shannon's rational inattention model where priors change across menus such that stability is broken even with full consideration.
\par 
We conclude by noting that the example in this section admits neither a RUM representation nor a RAM representation. However, it admits a set-monotone and stable RAUM representation. 

\section{Characterization of Set-monotone and Stable RAUM}\label{sec: characterization}
In this section, we characterize set-monotone and stable RAUM in a form amenable to (statistical) testing. In particular, we show that to conclude whether a given data set admits a set-monotone and stable RAUM representation, it suffices to check whether a particular linear program has a solution. This problem is similar to the one in \citet{mcfadden1990stochastic} that characterizes RUM. 
\par
First, we informally describe how to construct the linear program that needs to be solved. Suppose we have some $\rho$ and some $\pi$ and want to check whether $\pi$ is a RAUM representation of $\rho$. First, we need to check whether for all menus $A$ and $a\in A$, $\rho_{A}(a)$ is a mixture of preference-filter-types such that $a$ is considered and is the best among those considered: 
\[
\rho_A(a)=\sum_{(\succ,\phi)\in U\times\Phi}\pi_A(\succ,\phi)\Char{a\in\phi(A),\:a\succ b,\:\forall b\in\phi(A)\setminus\{a\}}.
\] 
If the choice set is $X=\{a,b\}$, preferences $U$ are such that $a\succ_1 b$ and $b\succ_2 a$, and 
there are only 3 feasible filters in $\Phi$: $\phi_1$, $\phi_2$ and $\phi_3$ such that $\phi_1(\{a\}) =\phi_1(\{a,b\})=\{a\}$,  $\phi_2(\{b\})=\phi_2(\{a,b\})=\{b\}$, $\phi_3(\{a,b\})=\{a,b\}$, then
\begin{align*}
    \pi_{\{a,b\}}(\succ_1,\phi_1)+\pi_{\{a,b\}}(\succ_1,\phi_3)+\pi_{\{a,b\}}(\succ_2,\phi_1)=\rho_{\{a,b\}}(a).
\end{align*}
Thus, iterating over $A$ and $a\in A$, we can construct a set of linear equality constrains that must be satisfied by $\pi$ and $\rho$.
\par
Second, the definition of RAUM requires $\pi_A(\succ,\phi)=0$ whenever $\phi(A)=\emptyset$, which is another linear equality constraint on $\pi$. For instance, $\pi_{\{b\}}(\succ_1,\phi_1)=0$.
\par
Third, each $\pi_A$ sums up to 1, giving us the third set of constraints. For instance,
\[
\pi_{\{a,b\}}(\succ_1,\phi_1)+\pi_{\{a,b\}}(\succ_1,\phi_2)+\pi_{\{a,b\}}(\succ_1,\phi_3)+\pi_{\{a,b\}}(\succ_2,\phi_1)+\pi_{\{a,b\}}(\succ_2,\phi_2)+\pi_{\{a,b\}}(\succ_2,\phi_3)=1.
\]
\par
Fourth, $\pi$ has to be stable. That is, $\sum_{\phi}(\pi_{A}(\succ,\phi)-\pi_{B}(\succ,\phi))=0$
for every $A,B\in\mathcal{A}$. For instance, 
\[
\pi_{\{a\}}(\succ_1,\phi_1)+\pi_{\{a\}}(\succ_1,\phi_2)-\pi_{\{a,b\}}(\succ_1,\phi_1)-\pi_{\{a,b\}}(\succ_1,\phi_2)=0.
\]
Thus, we get another set of linear constraints on $\pi$.
\par
Finally, to check set-monotonicity note that for $A\subseteq B$, under stability, $\pi_{A}(\phi|\succ)\geq\pi_{B}(\phi|\succ)$
is equivalent to 
\[
\pi_{A}(\phi|\succ)\pi^*(\succ)=\pi_{A}(\succ,\phi)\geq\pi_{B}(\succ,\phi)=\pi_{B}(\phi|\succ)\pi^*(\succ).
\]
To turn the inequality into equality, we can introduce a slack variable $\bar{v}_{A,B,\succ,\phi}\geq0$. As a result, we get that the final set of equality constraints is of the form
\[
\pi_{A}(\succ,\phi)-\pi_{B}(\succ,\phi)-\bar{v}_{A,B,\succ,\phi}=0.
\]
For example,
\[
\pi_{\{a\}}(\succ_1,\phi_1)-\pi_{\{a,b\}}(\succ_1,\phi_1)-\bar{v}_{\{a\},\{a,b\},\succ_1,\phi_1}=0.
\]
These are 5 types of linear equality restrictions on $v=(\pi',\bar{v}')'$, where $\bar{v}=(\bar{v}_{A,B,\succ,\phi})_{A\subseteq B\in\mathcal{A},\succ\in U,\phi\in \Phi}$ ($\bar{v}$ enters the first 4 restrictions with zero coefficients). Combining these linear equality restrictions for all menus, choices, filters, and preference orders, we can construct a matrix $G$ that consists of $0$, $1$, $-1$, and does not depend on $\rho$, and a vector $g$ that depends on $\rho$. In order to check whether $\pi$ is a RAUM representation of $\rho$ it is sufficient to check whether $g=Gv$. Our main theorem formalizes the above construction.
\par
Let $d$, $d_\rho$, $d_m$, and $d_r$ denote the length of $v$, the number of entries in $\rho$, the cardinality of $2^X\setminus\{\emptyset\}$, and the total number of linear restrictions-- imposed by feasibility, stability, and  set-monotonicity--on $\pi$, respectively. Also, define $g=(\rho\tr,1_{d_m}\tr,0_{d_r}\tr)\tr\in \Real^{d_g}$, where $1_{d_m}$ is the vector of ones of length $d_m$; $0_{d_r}$ is the vector of zeros of length $d_r$; and $d_g=d_{\rho}+d_m+d_r$.

\begin{theorem}\label{thm: Raum characterization}
Given a stochastic choice data set $\rho$ the following are equivalent:
\begin{enumerate}
\item $\rho$ admits a set-monotone and stable RAUM. 
\item There exists $v\in\Real^d_{+}$, $d<\infty$, such that 
\[
g=Gv
\]
where $G$ is a known matrix that consists of $-1$, $0$, $1$. 
\end{enumerate}
\end{theorem}
Theorem~\ref{thm: Raum characterization} provides a linear characterization of a set-monotone and stable RAUM. It is important to note that without stability the problem is quadratic since set-monotonicity is imposed on the conditional distribution over filters $\pi_A(\phi|\succ)$. The linearity of our problem makes it amenable to statistical testing using tools in \citet{deb2018revealed} as we discuss below.

\subsection*{Computational Aspects of Testing}
We have assumed that we observe $\rho$. In reality, we have to estimate $\rho$ from a sample of choices. To do this, in our preferred interpretation of RAUM, we need a cross-section of choices of a population of DMs with choice set variation.\footnote{One example of such data set is the one collected in \citet{aguiar2018does}. In that particular sample, there are $4099$ independent choices from a choice set with $|X|=6$ alternatives and $|\mathcal{A}|=32$ randomly and exogenously assigned menus.} 
Once we have the estimator of $\rho$, $\hat{\rho}$, we can use the testing procedure delineated in \citet{deb2018revealed}, which amounts to solve a (convex)-quadratic problem with linear constraints. Such problems are well-known in the optimization literature and typically easy to solve \citep{kitamura2018nonparametric,deb2018revealed}. The main computational cost can arise in the computation of the matrix $G$. The matrix $G$ does not depend on $\hat{\rho}_n$, hence, can be computed once and used for different datasets. Its size, however, grows exponentially with the size of the choice set $X$. For instance, for $|X|=6$, $G$ has about $1.8$ million rows and $4.4$ million columns. Fortunately, $G$ is sparse--for $|X|=6$ less than $0.00012$ percent of entries of $G$ are nonzero.\footnote{In our simulations, the computation time grows exponentially as well. While for $|X|=3$ matrix $G$ is computed in approximately $3\cdot 10^{-4}$ seconds, for $|X|=6$ it takes about $74$ minutes.} There are dedicated algorithms that can handle large-scale problems like ours by exploiting sparsity \citep[see, for instance,][]{benson2000solving,andersen2003implementing,goldfarb2005product,majumdar2020recent}.\footnote{Many of these recently proposed algorithms exploit the fact that quadratic and linear programs can be rewritten as semidefinite programming problems.}
\par 
In the extreme case where the choice set has a continuum of alternatives, such as in the framework of \citet{kitamura2018nonparametric}, the construction of the analogue matrix for the special case of RUM can become computational prohibitive. However, this level of complexity does not typically arise in our setup, as we focus on discrete choice with a moderate choice set size.
Another key difference from \citet{cattaneo2017random} is that we do not require the data set to be complete. We show that verifying the conditions on Theorem~\ref{thm: Raum characterization} is necessary and sufficient to guarantee that there is a set-monotone and stable RAUM representation of the data set. This means that our methodology can be applied to data sets that do not contain full variation in menus, such as those collected in \citet{apesteguia2018separating}.\footnote{Incompleteness of the data set also leads to substantial decrease in the size of $G$ since it is determined by $|\mathcal{A}|$.}

\section{Partial Identification of Preferences and Welfare Analysis}\label{sec: conditions}
\subsection*{Partial Identification of Preferences}
Although RAUM is falsifiable, given that preferences are not homogeneous, it is important to learn whether RAUM reveals anything about preferences. In this section, we show that RAUM reveals information about the distribution of preferences in population in some data sets. 
\par
We say that $\rho$ is regular if $\rho_A(a)\geq\rho_B(a)$ for all $A\subseteq B$ and $a\in A$. Otherwise, we call $\rho$ irregular. 
\par
To formalize the notion of revelation of preferences, let $\mathrm{R}_{\rho}$ be a set of all set-monotone and stable RAUM representations of $\rho$. That is, 
\[
\mathrm{R}_{\rho}=\left\{\pi\in \Delta(U\times\Phi)\::\: \pi \text{ is set-monotone and stable RAUM rule and } \rho \text{ admits } \pi \right\}.
\]
Next define the identified set for preference distributions implied by $\mathrm{R}_{\rho}$ as
\[
\Pi(\mathrm{R}_{\rho})=\left\{\pi^*\in\Delta(U)\::\:  \text{exists } \pi\in\mathrm{R}_{\rho}\text{ such that } \pi^*(\succ)=\sum_{\phi}\pi_A(\succ,\phi)\text{ for all } A,\succ \right\}.
\]
\begin{proposition}\label{prop: raum reveleation}
$\Pi(\mathrm{R}_{\rho})$ is a strict subset of $\Delta(U)$ for any irregular $\rho$. In particular, $\pi^*(\succ)<1$ for any $\succ\in U$ such that $b\succ a$ for all $b\in X$ and some $a$ in
\[
\{a\in X\::\:\rho_{B}(a)>\rho_{A}(a),\: a\in A\subseteq B,\:A,B\in\mathcal{A}\}.
\]
\end{proposition}
Proposition~\ref{prop: raum reveleation} states that irregular data is always informative about preferences. For example, if $\rho_{\{a,b,c\}}(a)>\rho_{\{a,b\}}(a)$, then we can conclude that $a$ cannot be the worst alternative with probability 1. There must exist a DM who ranks $a$ above something else. Since set-monotone and stable RAUM is a generalization of RAM, the conclusion of Proposition~\ref{prop: raum reveleation} is a generalization of the results in \citet{cattaneo2017random} for heterogeneous preferences.\footnote{Note that $\Pi(\mathrm{R}_{\rho})$ and $\Delta(U)$ are closed sets, hence, the difference between $\Delta(U)$ and $\Pi(\mathrm{R}_{\rho})$ has a positive Lebesgue measure.} 
\par
We conclude this section by noting that, since RAUM is a strict generalization of RAM, in general, the distribution over preferences can not be pined down uniquely without imposing more restrictions. Moreover, if there are several different preference orders that can explain the observed data set under RAM, then any distribution over these orders can explain the data set under RAUM. In particular, for regular data sets, nothing can be learned about preferences under RAUM, since there is no revelation of information about preferences under RAM \citep{cattaneo2017random}.

\subsection*{Out-of-Sample Predictions and Counterfactual Analysis}
Similar to \citet{kitamura2019nonparametric} who analyzed RUM, we can use our framework to conduct out-of-sample predictions and counterfactual analysis within the stable and set-monotone RAUM framework. In particular, we are interested in (i) predicting the choices of DMs in menus that are not observed in the data; and (ii) measuring welfare losses due to inattention as the fraction of individuals that do not achieve their first best due to limited consideration (i.e. the first best is the counterfactual situation where DMs consider the whole menu instead of its subsets).
\par
By Theorem~\ref{thm: Raum characterization}, we know that a given data set $\rho$, defined on the collection of menus $\mathcal{A}$, admits a set-monotone and stable RAUM representation $\pi$ if and only if there exists $v\in \Real_{+}^d$ such that
\begin{equation}\label{eq: pi set}
    g=Gv.
\end{equation}
Recall that if $v$ is a solution to Equation~(\ref{eq: pi set}), then the first $|U|\cdot(2^{|X|}-1)\cdot|\Phi|$ components of $v$ correspond to the set-monotone and stable RAUM rule. We will abuse notation and use $\pi^v$ to denote this rule. This $\pi^v$ includes rules for all menus (even those that are not observed in the data). Thus, the set of solutions to Equation~(\ref{eq: pi set}) characterizes all possible set-monotone and stable RAUM rules that can be admitted by a given $\rho$. 
\par
To make out-of-sample predictions, note that given $v$ (hence, $\pi^v$) that solves Equation~(\ref{eq: pi set}), $\pi^v_{A}(\succ,\phi)$ will be a fraction of DMs in the population with preferences $\succ$ and filter $\phi$ for any $A$ that is consistent with $\rho$. So we can compute the maximal out-of-sample probability of observing $a$ from $A\not\in\mathcal{A}$ (and $A\subseteq X$) that was not observed in $\rho$ as
\begin{align*}
    &\max_{\pi^v} \sum_{(\succ,\phi)\in U\times\Phi}\pi^v_A(\succ,\phi)\Char{a\in\phi(A),\:a\succ b,\:\forall b\in\phi(A)\setminus\{a\}},\\
    &\text{s.t. }Gv=g.
\end{align*}
The minimal out-of-sample probability of observing $a$ from $A\not\in\mathcal{A}$ (and $A\subseteq X$) can be computed in the same way by replacing the $\max$ by the $\min$ operator. 
\par
Next, to measure welfare losses due to inattention as the fraction of individuals that do not achieve their first best due to limited consideration,  for any $A\in\mathcal{A}$ such that $\sum_{\succ'}\pi_A(\succ',\phi)\neq 0$ define 
\begin{align*}
\Delta P_{A}(\pi^v)=&\sum_{a\in A,\phi\in\Phi,\succ\in U}\pi^v_A(\succ,\phi)\Big[\Char{a\in A,\:a\succ b,\:\forall b\in A\setminus\{a\}}\\
&-\Char{a\in \phi(A),\:a\succ b,\:\forall b\in \phi(A)\setminus\{a\}}\Big].
\end{align*}

$\Delta P_{A}(\pi^v)$ measures that counterfactual fraction of DMs endowed with menu $A$ who are strictly better off from considering all available alternatives in menu $A$. Hence, 
\begin{align*}
    &\max_{v\in\Real^d_{+}}\Delta P_{A}(\pi^v)\\
    &\text{s.t. }Gv=g
\end{align*}
will be the maximal fraction of DMs, who faced menu $A$, that would be better off if they consider all alternatives in $A$. If one is interested in the total effect it suffices to replace $\Delta P_{A}(v)$ by $\sum_{A\in\mathcal{A}}\Delta P_{A}(v)$ in the last optimization. Similar to the out-of-sample predictions, the lower bound of $\Delta P_{A}(\pi^v)$ can be computed by replacing the $\max$ by the $\min$ operator.
\par
Note that the counterfactual fully attentive behavior, assuming that each DM is fully attentive or considers the whole menu, is equivalent to the DMs behaving consistently with RUM governed by the true preference distribution (under our preferred interpretation of RAUM) $\pi^{*,v}$, where $\pi^{*,v}(\succ)=\sum_{\phi}\pi^v(\succ,\phi)$. 
\par
We highlight that our ordinal approach puts no restriction on the random utility distribution. We do not need to integrate over (unknown) distributions of parameters of high order polynomial approximations of the utility function or the consideration probability. Instead, by taking a purely revealed preference approach, our out-of-sample predictions and counterfactual welfare analysis require solving a linear program that delivers sharp bounds (i.e., a point is within bounds if and only if there exists a data generating process that is consistent with observed data and the point). 
\par 
We finish this section by remaking that $U$ can be restricted to any subset of linear orders exhibiting some property (e.g., single-crossing, \citealp{apesteguia2017single}, or expected utility, \citealp{kashaev2021random}). Our theory applies to these restrictions without changes. These restrictions, when valid, can improve the informativeness of the bounds studied in this section. However, we present our results for the unrestricted $U$ to maximize generality.

\section{Conclusions}\label{sec: conclusion}
We have extended the classical stochastic revealed preference methodology in \citet{mcfadden1990stochastic} for finite sets to allow for limited consideration. Our model allows for heterogeneous preferences that are correlated with consideration sets. We assume that consideration satisfies the set-monotonicity assumption of \citet{cattaneo2017random}.
We also introduce a new condition, called stability, that requires the marginal distribution of preferences to be independent of menus. We show that this new restriction is satisfied in many theoretical and empirical settings.
The proposed model and conditions are amenable to statistical testing using the procedure proposed in \citet{deb2018revealed}.

\bibliographystyle{econ}
\phantomsection\addcontentsline{toc}{section}{\refname}\bibliography{fieldconsideration}
\appendix

\section{Proofs}\label{app: proofs}
\subsection{Proof of Proposition~\ref{prop: emp bite of st and mon}}\label{app: emp bite of st and mon}
Since any $\rho$ can be completed, it is sufficient to establish validity of statements (i) and (ii) for complete stochastic data sets (i.e., $\mathcal{A}=2^X\setminus\{\emptyset\}$).
\par
\emph{Proof of (i).} Fix any complete $\rho$ and let $\pi_A(\phi|\succ)=\rho_A(a)\Char{\phi(A)=\{a\}}$ for all $a,A,\succ$. Then $\rho$ admits a stable RAUM representation $\pi_A(\phi|\succ)\pi^*(\succ)$, where $\pi^*$ is any element in $\Delta(U)$.
\par
\emph{Proof of (ii).} Fix any complete $\rho$. For any $\succ$ and $A$ let $a_{\succ,A}$ be the best element in $A$ according to $\succ$ and $\kappa_{\succ,A}=\sum_{\succ'\in U}\Char{a_{\succ,A}=a_{\succ',A}}$ be the number of preference orders for which $a_{\succ,A}$ is also the best. Take $\pi^*_A(\succ)=\rho_A(a_{\succ,A})/\kappa_{\succ,A}$. Then $\rho$ admits a set-monotone RAUM representation $\pi_A(\phi|\succ)\pi^*_A(\succ)$, where $\pi_{A}(\phi|\succ)=\Char{\phi(A)=A}$ for all $\succ$ and $A$.
\par
\emph{Proof of (iii).} To prove (iii) we will construct an incomplete data set (i.e., $\mathcal{A}\neq 2^X\setminus\{\emptyset\}$) that does not admit a set-monotone and stable RAUM.
Let $X=\{a,b,c,d\}$ and
\[
\mathcal{A}=\{\{a,b\}, \{a,c\},\{b,d\}, \{a,b,d\},\{a,c,d\}, \{b,c,d\}\}.
\]
Suppose the observed $\rho$ is as follows
\begin{equation*}
\begin{tabular}{c|c|c|c|c|c|c|}
\multicolumn{1}{c}{Menu}&\multicolumn{1}{c}{$\{a,b\}$}&\multicolumn{1}{c}{$\{a,c\}$}&\multicolumn{1}{c}{$\{b,d\}$}&\multicolumn{1}{c}{$\{a,b,d\}$}&\multicolumn{1}{c}{$\{a,c,d\}$}&\multicolumn{1}{c}{$\{b,c,d\}$}\\
\cline{2-7}
$\rho_A(a)$ & $1$ & $1$ & - & $0$ & $0$& -\\ 
\cline{2-7}
$\rho_A(b)$ & $0$ & - & $1$ & $1$ & -& $\alpha_b$\\ 
\cline{2-7}
$\rho_A(c)$ & - & $0$ & - & - & $1$& $\alpha_c$\\ 
\cline{2-7}
$\rho_A(d)$ & - & - & $0$ & $0$ & $0$& $\alpha_d$\\ 
\cline{2-7}
\end{tabular}
\end{equation*}
where $\alpha_d>0$. (Columns in the above matrix correspond to different menus. For instance, the third element of the second row is $\rho_{\{b,d\}}(b)$.) By way of contradiction assume that $\rho$ admits a set-monotone and stable RAUM.
We will abuse notation and associate filters with consideration sets they imply. For example, if $\phi_1$ is such that $\phi_1(A)=B$ we will write $\pi_{A}(\succ,B)$ instead of $\pi_{A}(\succ,\phi)$. Consider menus $\{a,b\}$ and $\{a,b,d\}$. Note that
\begin{align*}
    0=\rho_{\{a,b\}}(b)&=\sum_{\succ}\pi_{\{a,b\}}(\succ,\{b\})+\pi_{\{a,b\}}(\succ,\{a,b\})\Char{b\succ a},\\
    1=\rho_{\{a,b,d\}}(b)&=\sum_{\succ}\pi_{\{a,b,d\}}(\succ,\{b\})+\pi_{\{a,b,d\}}(\succ,\{a,b\})\Char{b\succ a}\\
    &+\pi_{\{a,b,d\}}(\succ,\{b,d\})\Char{b\succ d}+\pi_{\{a,b,d\}}(\succ,\{a,b,d\})\Char{b\succ a,d}.
\end{align*}
Subtracting the first equation from the second one, we get that
\begin{align*}
1&=\sum_{\succ}[\pi_{\{a,b,d\}}(\succ,\{b\})-\pi_{\{a,b\}}(\succ,\{b\})]+[\pi_{\{a,b,d\}}(\succ,\{a,b\})-\pi_{\{a,b\}}(\succ,\{a,b\})]\Char{b\succ a}\\
    &+\pi_{\{a,b,d\}}(\succ,\{b,d\})\Char{b\succ d}+\pi_{\{a,b,d\}}(\succ,\{a,b,d\})\Char{b\succ a,d}.
\end{align*}
Set-monotonicity of $\pi_{A}$ and stability of preferences then imply that
\begin{align*}
1&\leq\sum_{\succ}\pi_{\{a,b,d\}}(\succ,\{b,d\})\Char{b\succ d}+\pi_{\{a,b,d\}}(\succ,\{a,b,d\})\Char{b\succ a,d}.
\end{align*}
Since $\sum_{\succ}\sum_{D\subseteq\{a,b,d\}}\pi_{\{a,b,d\}}(\succ,D)=1$, we can conclude that the distribution over preferences $\pi^*$ is such that $b\succ d$ with probability 1. If we apply the above arguments to $\rho_{\{a,c\}}(c)$ and $\rho_{\{a,c,d\}}(c)$, we can deduce that $c\succ d$ with probability 1. Thus, with probability 1, $d$ is never picked if it is considered together with $b$ or $c$. Hence, in menu $\{b,c,d\}$ it can be picked with positive probability (i.e. $\alpha_d>0$) if and only if set $\{d\}$ is considered with positive probability. The later is not possible since $\pi_{\{b,c,d\}}(\succ,\{d\})\leq \pi_{\{b,d\}}(\succ,\{d\})\leq \rho_{\{b,d\}}(d)=0$ ($d$ is never picked in menu $\{b,d\}$). The contradiction completes the proof.

\subsection{Proof of Theorem~\ref{thm: Raum characterization}}\label{app: Raum characterization}
Assume that $\pi$ is a set-monotone and stable RAUM representation of possibly incomplete $\rho$. Let $d_m=\abs{2^X\setminus\{\emptyset\}\times U\times\Phi}$ and $\mathcal{A}_{a}=\{(a,A)\in X\times\mathcal{A}\::\:a\in A\}$. Fix any one-to-one mapping $i_{1}:\mathcal{A}_{a}\to\{1,2,\dots,\abs{\mathcal{A}_{a}}\}$ that maps a pair $(a,A)$ to a corresponding element of vector $\rho$. Also fix any one-to-one $i_{2}:2^X\setminus\{\emptyset\}\times U\times\Phi\to\{1,2,\dots,d_m\}$. Let $B$ be a matrix of size $\abs{\mathcal{A}_{a}}\times d_m$ such that the $(k,l)$-element of it, $B_{k,l}$, is defined as follows 
\[
B_{k,l}=\begin{cases}
\Char{a\succ b,\forall b\in\phi(A)\setminus\{a\}}, & \text{if }k=i_{1} ((a,A)),l=i_{2}(A,\succ,\phi)\text{ for some }\succ,\phi,(a,A)\in\mathcal{A}_a\\
0, & \text{otherwise}
\end{cases}.
\]
Hence, in matrix notation, if $\rho$ admits a RAUM representation, then 
\[
\rho=B\pi,
\]
where $\pi=(\pi_{A}(\succ,\phi))_{i_{2}(A,\succ,\phi)}$. 
\par 
The rest of the restrictions will be imposed on \emph{all} menus (including the ones that are not present in $\mathcal{A}$). These restrictions do not use any data.
First, we want to capture the fact that $\pi_A(\cdot,\cdot)$ is a probability distribution and needs to sum up to 1. For any $A\in 2^X\setminus\{\emptyset\}$, let an $i_{2}(A,\succ,\phi)$ element of a row of matrix $O$ to be 1 for all $\succ$ and $\phi$ and to be zero otherwise. Hence, the constraint can be written as 
\[
O\pi=1_{d_m},
\]
where $O$ is the matrix of size $d_m\times d_1$.
\par
The next set of restrictions captures feasibility: $\pi_A(\succ,\phi)=0$ whenever $\phi(A)=\emptyset$. Let $d_2=\sum_{A,\succ,\phi}\Char{\phi(A)=\emptyset}$. Then the feasibility constraint can be written as 
\[
F\pi=0,
\]
where $F$ is a matrix of 0/1 that picks $i_2(A,\succ,\phi)$ elements of $\pi$ that should be set to zero because of feasibility.
\par
Next we want to rewrite the definition of stability in the matrix form. Note that stability can be written as $\sum_{\phi}\pi_{A}(\succ,\phi)=\sum_{\phi}\pi_{B}(\succ,\phi)$
for all $A,B$. Fix any $A,B$, and $\succ$. Let $\iota^{A,B,\succ}$ be a vector of length $d$ such that 
\[
\iota_{k}^{A,B,\succ}=\Char{\exists\phi\::\:k=i_{2}(A,\succ,\phi)}-\Char{\exists\phi\::\:k=i_{2}(B,\succ,\phi)}.
\]
Take a collection of vectors $\left\{ \iota_{k}^{A,B,\succ}\right\} _{A,B,\succ}$ and remove all linearly dependent or zero vectors. Let every element of what is left to be a row of a matrix $S$. Then, stability is equivalent to 
\[
S\pi=0.
\]
\par
Finally, we want to build a matrix representation of set-monotonicity. Note that, under stability, $\pi_{A}(\phi|\succ)\geq\pi_{B}(\phi|\succ)$ is equivalent
to $\pi_{A}(\succ,\phi)\geq\pi_{B}(\succ,\phi)$. Hence, similarly to stability, fix any $A,B,\succ,\phi$ such that $A\subseteq B$, $A\neq B$, and let $\iota^{A,B,\succ,\phi}$ be
a vector of length $d$ such that 
\[
\iota_{k}^{A,B,\succ,\phi}=\Char{k=i_{2}(A,\succ,\phi)}-\Char{k=i_{2}(B,\succ,\phi)}.
\]
Similarly to matrix $S$ we can use vectors $\left\{ \iota_{k}^{A,B,\succ,\phi}\right\}$ to build matrix $M$ such that set-monotonicity is equivalent to 
\[
M\pi=\bar{v},
\]
where $\bar{v}$ is a component-wise nonnegative vector.
Define $G$ as 
\[
G=\left[\begin{array}{cc}
B & 0\\
O & 0\\
F & 0\\
S & 0\\
M & -I
\end{array}\right].
\]
As a result, if $\rho$ admits a set-monotone and stable RAUM representation, then the system $g=Gv$ has a component-wise nonnegative solution $(\pi\tr,\bar{v}\tr)\tr$. 
\par
Now suppose $g=Gv$ has a component-wise nonnegative solution $(\pi\tr,\bar{v}\tr)\tr$, we want to show that this $\pi$ is a set-monotone and stable RAUM representation of $\rho$. By the definition of $G$, $\pi$ is a complete (i.e., includes all possible menus) collection of distributions over $U\times\Phi$. Moreover, the constructed $\pi$ is set-monotone and stable and can generate the observed $\rho$. 

\subsection{Proof of Proposition~\ref{prop: raum reveleation}}
Towards a contradiction assume $\Pi_{\rho}=\Delta(U)$. If $\rho$ is irregular, then there exist $A,B\in\mathcal{A}$, $A\subseteq B$, and $a\in A$ such that $\epsilon\equiv\rho_B(a)-\rho_A(a)>0$. Since by assumption $\Pi_{\rho}=\Delta(U)$, take any $\pi$ such that $a$ is the worst with probability 1. If $\phi^*$ is such that $\phi^*(A)=\phi^*(B)=\{a\}$, then
\begin{align*}
    \rho_{B}(a)=\sum_{\succ}\sum_{\phi}\pi_B(\phi|\succ)\pi^*(\succ)\Char{a\succ b,\: \forall b\in \phi(B),\:a\in\phi(B)}=\sum_{\succ}\pi_{B}(\phi^*|\succ)\pi^*(\succ).
\end{align*}
Similarly,
\[
    \rho_{A}(a)=\sum_{\succ}\pi_{A}(\phi^*|\succ)\pi^*(\succ).
\]
Taking the difference between these two equations we get that
\begin{align*}
    0<\epsilon&=\sum_{\succ}[\pi_{B}(\phi^*|\succ)-\pi_{A}(\phi^*|\succ)]\pi^*(\succ)\leq 0,
\end{align*}
where the last inequality follows from set-monotonicity. This contradiction completes the proof.

\section{Omitted Details from Example~\ref{ex: RI}}\label{app: RI1}
There are 4 possible non-singleton menus: $\{a,b,c\}$, $\{a,b\}$, $\{b,c\}$, and $\{a,c\}$. Given $\delta$ and $A$, let $D_{\delta,A}$ be the deterministic consideration set. 
\par
\noindent \textbf{Case 1, $A=\{a,b,c\}$.} Theorem~1 in \citet{caplin2016rationalconsideration} implies that if $\mu(c)>1/(3+\delta)$ (i.e., $\delta>-3+1/\mu(c)=\delta^*_1$), then $D_{\delta,A}=A$. If $\delta<\delta^*_1$ and
\[
\mu(b)\geq\dfrac{\mu(a)+\mu(b)}{2+\delta}>\mu(c)
\]
(i.e., $\delta_4^*\leq\delta<\delta_1^*$), then $D_{\delta,A}=\{a,b\}$. Finally, if 
\[
\mu(a)\geq\dfrac{\mu(a)}{1+\delta}>\mu(b)
\]
(i.e., $\delta<\delta_4^*$), then the consideration set is $\{a\}$.
\par
\noindent \textbf{Case 2, $A=\{a,b\}$.} Applying the same Theorem~1 in \citet{caplin2016rationalconsideration} we can obtain that if
\[
\mu_A(b)=\dfrac{\mu(b)}{\mu(a)+\mu(b)}>1/(2+\delta)
\]
(i.e., $\delta>-1+\mu(a)/\mu(b)=\delta_4^*$), then $D_{\delta,A}=\{a,b\}$, and $D_{\delta,A}=\{a\}$ otherwise.
\par
\noindent \textbf{Case 3, $A=\{b,c\}$.} Similarly to Case 2, if
\[
\mu_A(b)=\dfrac{\mu(c)}{\mu(b)+\mu(c)}>1/(2+\delta)
\]
(i.e., $\delta>-1+\mu(b)/\mu(c)=\delta_e^*$), then $D_{\delta,A}=\{b,c\}$, and $D_{\delta,A}=\{b\}$ otherwise.
\par
\noindent \textbf{Case 4, $A=\{a,c\}$.} If
\[
\mu_A(c)=\dfrac{\mu(c)}{\mu(a)+\mu(c)}>1/(2+\delta)
\]
(i.e., $\delta>-1+\mu(a)/\mu(c)=\delta_2^*$), then $D_{\delta,A}=\{a,c\}$, and $D_{\delta,A}=\{a\}$ otherwise.
\par
Given that $1-2\mu(c)=\mu(a)+(\mu(b)-\mu(c)\geq\mu(a)$ and $\mu(a)\geq\mu(b)\geq\mu(c)$, it follows that $\delta^*_1\geq\delta^*_2\geq\delta^*_3$ and $\delta^*_2\geq\delta^*_4$. 
Tables~\ref{tab:RI1} and~\ref{tab:RI2} summarize the above derivations for different values of $\delta$ and menu $A$.  Table~\ref{tab:RI1} corresponds to the case when $\delta_3^*\geq \delta_4^*$. Table~\ref{tab:RI2} considers the case when $\delta_4^*<\delta_3^*$.   
\par
Next, we compute the implied by the model probabilities of choosing different options. Theorem~1 in \citet{caplin2016rationalconsideration} implies that the probability that $y\in D_{\delta,A}$ is chosen from $y\in D_{\delta,A}$ satisfies
\[
P_{D_{\delta,A}}(y)=\dfrac{\mu_{D_{\delta,A}}(y(|D_{\delta,A}|+\delta)-1)}{\delta}.
\]
Assume that $\delta_3^*>\delta_4^*$ (the opposite case leads to the same conclusion). The following table displays $P_{D_{\delta,A}}(a)$. 
\begin{equation*}
\begin{tabular}{c|c|c|c|c|c|}
\multicolumn{1}{c}{$A/\delta\in$}&\multicolumn{1}{c}{$(0,\delta^*_4]$}&\multicolumn{1}{c}{$(\delta^*_4,\delta^*_3]$}&\multicolumn{1}{c}{$(\delta^*_3,\delta^*_2]$}&\multicolumn{1}{c}{$(\delta^*_2,\delta^*_1]$}&\multicolumn{1}{c}{$(\delta^*_1,+\infty)$}\\
\cline{2-6}
$\{a,b,c\}$ & $1$ & $P_{\{a,b\}}(a)$ & $P_{\{a,b\}}(a)$ & $P_{\{a,b\}}(a)$ & $P_{\{a,b,c\}}(a)$\\ 
\cline{2-6}
$\{a,b\}$ & 1 & $P_{\{a,b\}}(a)$ & $P_{\{a,b\}}(a)$ & $P_{\{a,b\}}(a)$ & $P_{\{a,b\}}(a)$\\ 
\cline{2-6}
$\{a,c\}$ & 1 & 1 & 1 & $P_{\{a,c\}}(a)$ & $P_{\{a,c\}}(a)$\\ 
\cline{2-6}
\end{tabular}
\end{equation*}
For options $b$ and $c$ the tables are
\begin{equation*}
\begin{tabular}{c|c|c|c|c|c|}
\multicolumn{1}{c}{$A/\delta\in$}&\multicolumn{1}{c}{$(0,\delta^*_4]$}&\multicolumn{1}{c}{$(\delta^*_4,\delta^*_3]$}&\multicolumn{1}{c}{$(\delta^*_3,\delta^*_2]$}&\multicolumn{1}{c}{$(\delta^*_2,\delta^*_1]$}&\multicolumn{1}{c}{$(\delta^*_1,+\infty)$}\\
\cline{2-6}
$\{a,b,c\}$ & $0$ & $P_{\{a,b\}}(b)$ & $P_{\{a,b\}}(b)$ & $P_{\{a,b\}}(b)$ & $P_{\{a,b,c\}}(b)$\\ 
\cline{2-6}
$\{a,b\}$ & 0 & $P_{\{a,b\}}(b)$ & $P_{\{a,b\}}(b)$ & $P_{\{a,b\}}(b)$ & $P_{\{a,b\}}(b)$\\ 
\cline{2-6}
$\{b,c\}$ & 1& 1 & $P_{\{b,c\}}(b)$ & $P_{\{b,c\}}(b)$ & $P_{\{b,c\}}(b)$\\ 
\cline{2-6}
\end{tabular}
\end{equation*}
and
\begin{equation*}
\begin{tabular}{c|c|c|c|c|c|}
\multicolumn{1}{c}{$A/\delta\in$}&\multicolumn{1}{c}{$(0,\delta^*_4]$}&\multicolumn{1}{c}{$(\delta^*_4,\delta^*_3]$}&\multicolumn{1}{c}{$(\delta^*_3,\delta^*_2]$}&\multicolumn{1}{c}{$(\delta^*_2,\delta^*_1]$}&\multicolumn{1}{c}{$(\delta^*_1,+\infty)$}\\
\cline{2-6}
$\{a,b,c\}$ & $0$ & 0 & 0 & 0 & $P_{\{a,b,c\}}(c)$\\ 
\cline{2-6}
$\{b,c\}$ & 0& 0 & $P_{\{b,c\}}(c)$ & $P_{\{b,c\}}(c)$ & $P_{\{b,c\}}(c)$\\ 
\cline{2-6}
$\{a,c\}$ & 0 & 0 & 0 & $P_{\{a,c\}}(c)$ & $P_{\{a,c\}}(c)$\\ 
\cline{2-6}
\end{tabular}
\end{equation*}
Note that since $P_{\{a,b,c\}}(a)< P_{\{a,b\}}(a)\leq P_{\{a,c\}}(a)$, $P_{\{a,b,c\}}(b)< P_{\{a,b\}}(b)\leq P_{\{b,c\}}(b)$, and $P_{\{a,b,c\}}(c)< P_{\{a,c\}}(c)\leq P_{\{b,c\}}(c)$, the computes distributions do not violate regularity for all values of $\delta$ and all menus $A$.
\par
If $\succ_{y_1y_2y_3}$ is such that $y_1\succ_{y_1y_2y_3} y_2\succ_{y_1y_2y_3} y_3$, then one can verify that the following $\pi^*$ is a stable distribution that together with the deterministic consideration set is consistent with observed choices:
\begin{align*}
    \pi^*(\succ_{abc})&=\dfrac{(\delta-\delta^*_1+\kappa_{ac})}{(\mu(b)+\mu(c))}\dfrac{\mu(a)\mu(b)}{\delta}\geq0,\\
    \pi^*(\succ_{acb})&=\dfrac{(\delta-\delta^*_1+\kappa_{ac})}{(\mu(b)+\mu(c))}\dfrac{\mu(a)\mu(c)}{\delta}\geq0,\\
    \pi^*(\succ_{bac})&=\dfrac{(\delta-\delta^*_1+\kappa_{bc})}{(\mu(a)+\mu(c))}\dfrac{\mu(a)\mu(b)}{\delta}\geq0,\\
    \pi^*(\succ_{bca})&=\dfrac{(\delta-\delta^*_1+\kappa_{bc})}{(\mu(a)+\mu(c))}\dfrac{\mu(b)\mu(c)}{\delta}\geq0,\\
    \pi^*(\succ_{cab})&=\dfrac{(\delta-\delta^*_1)}{(\mu(a)+\mu(b))}\dfrac{\mu(a)\mu(c)}{\delta}\geq0,\\
    \pi^*(\succ_{cba})&=1-(\pi_1+\pi_2+\pi_3+\pi_4+\pi_5)=\dfrac{(\delta-\delta^*_1)}{(\mu(a)+\mu(b))}\dfrac{\mu(b)\mu(c)}{\delta}\geq0,
\end{align*}
where $\kappa_{ac}=1/\mu(c)-1/\mu(a)\geq 0$ and $\kappa_{bc}=1/\mu(c)-1/\mu(b)\geq 0$.

\section{Omitted Details from Section~\ref{sec: examples}}\label{app: gen formula}
In this appendix, we verify the general formula for the attention-index models in Example~\ref{ex: Attentionindex} for the representation of the logit attention model \citep{brady2016menu} and the elimination by aspects \citep{tversky1972elimination,aguiar2017random}. Recall that 
\[
\pi_A(\phi|\succ)=\dfrac{\sum_{B\subseteq X\setminus \tilde{g}(A)}\eta_{\succ}(\phi(A)\cup B)}{\sum_{C\in 2^A\setminus\{\emptyset\}}\sum_{B\subseteq X\setminus \tilde{g}(A)}\eta_{\succ}(C\cup B)},
\]
for a known $\tilde{g}:2^X\to2^X$.
\par
To prove the relation for the logit attention model, note that if $\tilde{g}(A)=X$ for all $A$, then the probability of considering $D=\phi(A)$ in menu $A$, $m_{\succ,A}(D)$, is equal to
\[
m_{\succ,A}(D)=\pi_A(\phi|\succ)=\dfrac{\sum_{B\subseteq X\setminus X}\eta_{\succ}(D\cup B)}{\sum_{C\in 2^A\setminus\{\emptyset\}}\sum_{B\subseteq X\setminus X}\eta_{\succ}(C\cup B)}=\dfrac{\eta_{\succ}(D)}{\sum_{C\in 2^A\setminus\{\emptyset\}}\eta_{\succ}(C)}.
\]
The latter corresponds to the consideration rule in \citet{brady2016menu}, since, by construction, $\eta_{\succ}(\emptyset)=0$.
\par
Similarly, by definition, the consideration rule of the elimination by aspects is
\[
m_{\succ,A}(D)=\sum_{C\subseteq X\::\:C\cap A=D}\frac{\eta_{\succ}(C)}{\sum_{K\subseteq X\::\:K\cap A\neq\emptyset}\eta_{\succ}(K)}.
\]
From 
\[
\sum_{K\::\:K\cap A\neq\emptyset}\eta_{\succ}(K)=\sum_{B\subseteq X\setminus A}\sum_{C\subseteq A,\:C\neq\emptyset}\eta_{\succ}(C\cup B)
\]
and 
\[
\sum_{C\::\:C\cap A=D}\eta_{\succ}(C)=\sum_{B\subseteq X\setminus A}\eta_{\succ}(D\cup B),
\]
it follows that
\begin{align*}
    m_{\succ,A}(D)=\dfrac{\sum_{B\subseteq X\setminus A}\eta_{\succ}(D\cup B)}{\sum_{C\in 2^A\setminus\{\emptyset\}}\sum_{B\subseteq X\setminus A}\eta_{\succ}(C\cup B)}.
\end{align*}
Hence, $\tilde{g}(A)=A$ generates the elimination by aspects consideration rule.
\par 
Now we establish a sufficient condition on the mapping $\tilde{g}$ that imply set-monotonicity on the induced $\pi$ rule. 
\begin{lemma}\label{lemma:conditionong}
If $\tilde{g}(A)\subseteq \tilde{g}(A')$ and $\tilde{g}(A')\setminus \tilde{g}(A)\subseteq A'\setminus A$ for all $A\subseteq A'$, then 
\[
\pi_A(\phi|\succ)=\dfrac{\sum_{B\subseteq X\setminus \tilde{g}(A)}\eta_{\succ}(\phi(A)\cup B)}{\sum_{C\in 2^A\setminus\{\emptyset\}}\sum_{B\subseteq X\setminus \tilde{g}(A)}\eta_{\succ}(C\cup B)},
\]
satisfies set-monotonicity.
\end{lemma}
This restriction on $\tilde{g}$ is not exhausted by the logit attention and the elimination by aspects models. For example, $\tilde{g}(A)=A\setminus\{a^*,b^*\}$ for all $A\in \mathcal{A}$; and  $\tilde{g}(A)=X\setminus\{a^*,b^*\}$ for all $A\in \mathcal{A}$, where $\{a^*,b^*\}\subseteq X$ are fixed items both satisfy the conditions of Lemma~\ref{lemma:conditionong}. The extent to which these new mappings $\tilde{g}$ induce an empirically relevant consideration rules is outside the scope of this paper. 

\subsection*{Proof of Lemma~\ref{lemma:conditionong}. }
\begin{proof}
Let $D=\phi(A)=\phi(A')$ for $A\subseteq A'$. First, note that
\begin{align}\label{eq: prp gen 1}
    \sum_{B\subseteq X\setminus \tilde{g}(A)}\eta_{\succ}(D\cup B)-\sum_{B\subseteq X\setminus \tilde{g}(A')}\eta_{\succ}(D\cup B) \geq 0   
\end{align}
since $X\setminus \tilde{g}(A') \subseteq X\setminus \tilde{g}(A)$. Second, let $L=A'\setminus A$ and $T=\tilde{g}(A')\setminus \tilde{g}(A)$, and note that
\begin{align*}
    &\sum_{C\in 2^A\setminus\{\emptyset\}}\sum_{B\subseteq X\setminus \tilde{g}(A)}\eta_{\succ}(C\cup B)=\sum_{C\in 2^A\setminus\{\emptyset\}}\sum_{B\subseteq X\setminus \tilde{g}(A')}\eta_{\succ}(C\cup B)+\sum_{C\in 2^A\setminus\{\emptyset\}}\sum_{B\subseteq X\setminus \tilde{g}(A')}\sum_{K\subseteq T,K\neq\emptyset}\eta_{\succ}(C\cup B\cup K),\\
    &\sum_{C\in 2^{A'}\setminus\{\emptyset\}}\sum_{B\subseteq X\setminus \tilde{g}(A')}\eta_{\succ}(C\cup B)=\sum_{C\in 2^A\setminus\{\emptyset\}}\sum_{B\subseteq X\setminus \tilde{g}(A')}\eta_{\succ}(C\cup B)+\sum_{C\in 2^A\setminus\{\emptyset\}}\sum_{B\subseteq X\setminus \tilde{g}(A')}\sum_{K\subseteq L,K\neq\emptyset}\eta_{\succ}(C\cup B\cup K).
\end{align*}
As a result, since we assume that $T\subseteq L$, we can conclude that
\begin{align}\label{eq: prp gen 2}
    \nonumber&\sum_{C\in 2^{A'}\setminus\{\emptyset\}}\sum_{B\subseteq X\setminus \tilde{g}(A')}\eta_{\succ}(C\cup B)-\sum_{C\in 2^{A}\setminus\{\emptyset\}}\sum_{B\subseteq X\setminus \tilde{g}(A)}\eta_{\succ}(C\cup B)=\\
    &\sum_{C\in 2^A\setminus\{\emptyset\}}\sum_{B\subseteq X\setminus \tilde{g}(A')}\sum_{K\subseteq L\setminus T,K\neq\emptyset}\eta_{\succ}(C\cup B\cup K) \geq 0   
\end{align}
Hence,
\begin{align*}
    \pi_A(\phi|\succ)-\pi_{A'}(\phi|\succ)&=\dfrac{\sum_{B\subseteq X\setminus \tilde{g}(A)}\eta_{\succ}(\phi(A)\cup B)}{\sum_{C\in 2^A\setminus\{\emptyset\}}\sum_{B\subseteq X\setminus \tilde{g}(A)}\eta_{\succ}(C\cup B)}-\dfrac{\sum_{B\subseteq X\setminus \tilde{g}(A')}\eta_{\succ}(\phi(A')\cup B)}{\sum_{C\in 2^{A'}\setminus\{\emptyset\}}\sum_{B\subseteq X\setminus \tilde{g}(A')}\eta_{\succ}(C\cup B)}\geq\\
    &\geq\dfrac{\sum_{B\subseteq X\setminus \tilde{g}(A)}\eta_{\succ}(\phi(A)\cup B)}{\sum_{C\in 2^A\setminus\{\emptyset\}}\sum_{B\subseteq X\setminus \tilde{g}(A)}\eta_{\succ}(C\cup B)}-\dfrac{\sum_{B\subseteq X\setminus \tilde{g}(A)}\eta_{\succ}(\phi(A)\cup B)}{\sum_{C\in 2^{A'}\setminus\{\emptyset\}}\sum_{B\subseteq X\setminus \tilde{g}(A')}\eta_{\succ}(C\cup B)}\geq\\
    &\geq\dfrac{\sum_{B\subseteq X\setminus \tilde{g}(A)}\eta_{\succ}(\phi(A)\cup B)}{\sum_{C\in 2^{A'}\setminus\{\emptyset\}}\sum_{B\subseteq X\setminus \tilde{g}(A')}\eta_{\succ}(C\cup B)}-\dfrac{\sum_{B\subseteq X\setminus \tilde{g}(A)}\eta_{\succ}(\phi(A)\cup B)}{\sum_{C\in 2^{A'}\setminus\{\emptyset\}}\sum_{B\subseteq X\setminus \tilde{g}(A')}\eta_{\succ}(C\cup B)}=0,
\end{align*}
where the first inequality follows from Equation (\ref{eq: prp gen 1}) and the second one follows from Equation (\ref{eq: prp gen 2}).
\end{proof}
\end{document}